\documentclass[fleqn,usenatbib]{mnras}

\usepackage{newtxtext,newtxmath}

\usepackage[T1]{fontenc}

\DeclareRobustCommand{\VAN}[3]{#2}
\let\VANthebibliography\thebibliography
\def\thebibliography{\DeclareRobustCommand{\VAN}[3]{##3}\VANthebibliography}

\usepackage{graphicx}	
\usepackage{amsmath}	
\usepackage{academicons} 
\usepackage{xcolor}
\usepackage{widetext}
\usepackage{url}
\usepackage[utf8]{inputenc}
\usepackage{hyperref}
\usepackage{orcidlink}
\usepackage{multicol}
\usepackage{pdflscape}

\newcommand{\msun}{M$_{\odot}~$} 
\newcommand{\msune}{M$_{\odot}$}

\newcommand{\dgr}{$^{\circ}~$}

\newcommand{\teff}{$\rm T_{eff}$~}
\newcommand{\teffe}{$\rm T_{eff}$}
\newcommand{\logg}{$\log{g}$~}
\newcommand{\logge}{$\log{g}$}

\title[LMC Ages]{Revealing the Chemical Structure of the Magellanic Clouds with APOGEE. I. Calculating Individual Stellar Ages of RGB Stars in the Large Magellanic Cloud}

\author[Povick et al.]{Joshua T. Povick\orcidlink{0000-0002-6553-7082}$^{1}$\thanks{E-mail: joshua.povick@montana.edu},
David L. Nidever\orcidlink{0000-0002-1793-3689}$^{1}$,
Pol Massana\orcidlink{0000-0002-8093-7471}$^{1}$,
Jamie Tayar\orcidlink{0000-0002-4818-7885}$^{2,3}$,
Knut A.G. Olsen\orcidlink{https://orcid.org/0000-0002-7134-8296}$^4$,
\newauthor
 Sten Hasselquist\orcidlink{0000-0001-5388-0994}$^{5,6}$,
Maria-Rosa L. Cioni\orcidlink{0000-0002-6797-696X}$^{7}$,
Christian Nitschelm\orcidlink{0000-0003-4752-4365}$^{8}$, 
Ricardo Carrera$^{9}$,
\newauthor
Yumi Choi$^{4,10}$,
Alexandre Roman-Lopes\orcidlink{0000-0002-1379-4204}$^{11}$,
Steven R. Majewski\orcidlink{0000-0003-2025-3147}$^{12}$,
Andr\'{e}s Almeida$^{12}$,
\newauthor
Katia Cunha\orcidlink{0000-0001-6476-0576}$^{13,14,15}$ and
Verne V. Smith\orcidlink{0000-0002-0134-2024}$^{4}$ \\
$^{1}$Department of Physics, Montana State University, P.O. Box 173840, Bozeman, MT 59717-3840\\
$^{2}$Department of Astronomy, University of Florida, Bryant Space Science Center, Stadium Road, Gainesville, FL 32611, USA \\
$^{3}$Institute for Astronomy, University of Hawai‘i at M\={a}noa, 2680 Woodlawn Drive, Honolulu, HI 96822, USA \\ 
$^{4}$NSF’s National Optical-Infrared Astronomy Research Laboratory, 950 N.
Cherry Ave., Tucson, AZ 85719, USA \\
$^{5}$Department of Physics \& Astronomy, University of Utah, Salt Lake City, UT 84112, USA \\
$^{6}$Space Telescope Science Institute, 3700 San Martin Drive, Baltimore, MD 21218 \\
$^{7}$Leibniz-Institut f\"{u}r Astrophysik Potsdam, An der Sternwarte 16, D-14482 Potsdam, Germany \\
$^{8}$Centro de Astronom{\'i}a (CITEVA), Universidad de Antofagasta, Avenida Angamos 601, Antofagasta 1270300, Chile \\
$^{9}$INAF-Osservatorio di Astrofisica e Scienza dello Spazio di Bologna, via P. Gobetti 93/3, 40129, Bologna, Italy \\
$^{10}$Department of Astronomy, University of California Berkeley, Berkeley, CA 94720, USA \\
$^{11}$Departamento de Astronomia, Facultad de Ciencias, Universidad de La Serena, Av. Juan Cisternas 1200, La Serena, Chile \\
$^{12}$Department of Astronomy, University of Virginia, Charlottesville, VA 22904-4325, USA \\
$^{13}$Steward Observatory, University of Arizona, 933 North Cherry Avenue, Tucson, AZ, 85721, USA \\
$^{14}$Observat\'{o}rio Nacional, Rua General Jos\'{e} Cristino, 77, 20921-400 S\~{a}o Crist\'{o}v\~{a}o, Rio de Janeiro, RJ, Brazil \\
$^{15}$Institut d'Astrophysique de Paris, UMR 7095 CNRS, Sorbonne Universit\'{e}, 98bis Bd. Arago, 75014 Paris, France
}

\date{Accepted XXX. Received YYY; in original form ZZZ}

\pubyear{2023}

\begin{document}
\label{firstpage}
\pagerange{\pageref{firstpage}--\pageref{lastpage}}
\maketitle

\begin{abstract}
Stellar ages are critical for understanding the temporal evolution of a galaxy. We calculate the ages of over 6000 red giant branch stars in the Large Magellanic Cloud (LMC) observed with SDSS-IV / APOGEE-S. Ages are derived using multi-band photometry, spectroscopic parameters (\teffe, \logge, [Fe/H], and [$\alpha$/Fe]) and stellar isochrones and the assumption that the stars lie in a thin inclined plane to get accurate distances.  The isochrone age and extinction are varied until a best match is found for the observed photometry. 
We perform validation using the APOKASC sample, which has asteroseismic masses and accurate ages, and find that our uncertainties are $\sim$20\% and range from $\sim$1$-$3 Gyr for the calculated age values.
Here we present the LMC age map as well as the age-radius relation and an accurate age-metallicity relation (AMR). The age map and age-radius relation reveal that recent star formation in the galaxy was more centrally located and that there is a slight dichotomy between the north and south with the northern fields being slightly younger. The northern fields that cover a known spiral arm have median ages of $\gtrsim$ 2 Gyr, which is the time when an interaction with the SMC is suggested to have happened. The AMR is mostly flat especially for older ages although recently (about 2.0--2.5 Gyr ago) there is an increase in the median [Fe/H]. Based on the time frame, this might also be attributed to the close interaction between the LMC and SMC.

\end{abstract}




\section{Introduction}
\label{sec:ages_intro}

The ages of stars is very important for understanding galaxy evolution and galactic archaeology.
Without temporal information it becomes much more difficult to pinpoint how galaxies evolved to what we see today. Unfortunately, determining ages is not without its challenges as we rely on indirect means to find ages. Often times limited photometry or low resolution spectra make ages nearly impossible to determine because there is not enough information to constrain the ages of stars provided by these methods. 



Stellar clusters have been accurately age-dated for many decades, taking 
advantage of the age-sensitivity of turnoff and subgiant branch stars \citep[e.g.][]{sandage1970msto,flower1983msto,sarajedini2008msto}.
It is substantially more challenging to measure the ages of individual field stars.
However, there are several methods that have been developed to calculate ages of individual stars.
Some of the more popular methods include: nucleocosmochronometry, gyrochronology, asteroseismology, chemical abundance ratios, and using isochrones. Nucleocosmochronometry relies on measuring the abundance of radioactive nuclides and their daughter products and is much like the analog of radiocarbon dating, but for stars \citep{hill2002first,frebel2007discovery}.  Very high signal-to-noise and high resolution spectra are required to make these measurements and even then the relevant absorption lines may not exist in the spectrum.  In general, nucleocosmochronometry has only been performed for a small number of stars.
Gyrochronology uses the rotation rate of a star to calculate its age because as a star ages its rotational speed decreases \citep{barnes2007ages, mamajek2008improved}.  This requires precise, time-series photometry of cool, main-sequence stars which can be challenging to obtain for external galaxies like the LMC.
Asteroseismology uses scaling relations and the oscillations inside of a star to obtain precise masses and then ages using isochrones \citep{cunha2007asteroseismology,pinsonneault2014apokasc}.  Since mass and age are strongly anticorrelated on the giant branch, measuring mass accurately also allows for an accurate age to be determined.
However, very precise and high-cadence photometry is required to derive good asterseismologic masses and this is currently only available for small portions of the sky (i.e., Kepler, K2, CoRoT, TESS). 
More recently, spectroscopic chemical abundances have been used to determine masses of RGB stars with [C/N] and [C$^{12}$/C$^{13}$] ratios \citep{ness2016}.  These ratios are sensitive to the mass-dependent dredge-up that happens on the giant branch and pulls up nuclear-processed material from the interior.  Currently, this technique only works for metal-rich populations ([Fe/H]$\gtrsim$ $-$0.5), and, therefore, is not applicable for the relatively metal-poor LMC stars.
Arguably one of the most popular methods to find ages is using isochrones. Isochrone-fitting involves fitting theoretical stellar models to either individual or groups of stars. Many of these methods rely on comparing to stellar evolutionary models and to be able to truly do this we need many photometric bands that constrain the spectral energy distribution (SED) or high resolution spectra.  The {\tt BEAST} \citep{gordon2016} compares a grid of isochrones to multi-band photometry to determine mass, age, metallicity, \teffe, and \logg on a star-by-star basis.  This has been used successfully in M31 using the PHAT survey \citep{dalcanton2012}.  
All of these methods come with their own set of pros and cons and the choice of which one to use is often dictated by context.




Previous work of finding ages using isochrones for giant stars include works such as \cite{feuillet2016ages}. In that work ages are found two different ways: 1) calculating the mass from measured stellar parameters from which the mass--age relation can be used to get age and 2) using a Bayesian isochrone matching technique with a constant star formation history (SFH) to derive an age probability distribution for each star. Through mock data tests it was found that this method shows that decent ages can be derived for individual stars using a combination of photometry and spectroscopy with isochrones using a probabilistic approach.


This work uses isochrones to determine the age of individual red giant branch (RGB) stars using photometric and spectroscopic observations through direct calculation that is non-probabilistic. We take advantage of the fact that the absolute magnitude of giant stars vary substantially with age for fixed \teffe, by as much as 0.8 magnitudes between 1 Gyr and 10 Gyr at an average LMC metallicity of [Fe/H]=$-$0.5.
One potential drawback of using isochrones is the age-metallicity degeneracy of RGB stars, but we overcome this by using the metallicity measured with high-resolution spectroscopy. 





The method presented in this work finds the age by calculating the model photometry  age in six different passbands spanning both optical and infrared wavelengths using a single common trial and comparing to the observed photometry of a star. This method is possible due to accurate distances, which are needed to constrain the absolute magnitude of a star.




There are two main reasons for studying the ages of red giant branch (RGB) stars in the LMC. The first reason is that the literature is scant on the spatial distribution of ages in the galaxy especially using individual stars. The LMC is a relatively close satellite galaxy of the Milky Way (MW) at 49.9 kpc \citep{van2001magellanic,degrijs2014clustering} and so it makes sense to analyze it as it does not pose as many challenges compared to galaxies which are farther away that may or may not be resolved.
The second reason is that accurate ages are needed on a star-to-star basis to study the chemical evolution of the LMC at very high resolution.

Galaxies are the sum of their parts and stars make up a large portion of them. With a large enough sample of stars with good spatial coverage, it becomes possible to study the chemical evolution of a galaxy. When a star ``dies'', material is ejected into the interstellar medium (ISM) making it available to form newer stars. Thus enriching the reservoir of gas in the ISM to higher metallicities through the yields of the now deceased star. This leads to new stars having elevated chemical abundances and hence higher metallicities. It is also well known that metallicity is a proxy for age. While the particular age-metallicity (AMR) for galaxies will differ, they are an extremely useful tool to study galaxy evolution. The AMR reveals how a galaxy evolved over time. Generally the AMR is measured using clusters or field stars.



Studies of the star formation history (SFH) of a galaxy have the potential to yield AMRs. Usually, observations of resolved stellar population observations are fit with stellar evolutionary models using a combination of distance, age and metallicity parameters. This means that as a result one ends up having a relation between the age and metallicity of stars in the galaxy. Normally though, this method gives very approximate results. For the LMC, there are very few examples where these results are used to draw conclusions about the galaxy's evolution. \cite{harris2009lmcsfh} used the MCPS survey \citep{zartisky1997} to derive SFH across the main-body of the LMC. Because of the shallow nature of the photometric survey, the study was limited to younger ages and only four metallicity bins per single age stellar population. The general conclusions were that the LMC seemed to remain at a constant metallicity in the old ages until around $\approx 4$ Gyr ago, when it started to increase to the present day value. A more recent approach with larger photometric depths was used by \citet{meschin2014}, covering three small fields in the LMC. The AMR derived from the SFH results had a good fit with spectroscopic results from field stars by \cite{carrera2008lmcchem}, but the spatial scope of the study was still relatively small. Furthermore, \citet{weisz2013} used deep $HST$ imaging to derive the SFH in eight fields in the LMC (as well as seven more in the SMC), but released no metallicity information due to the focus being on the comparison of star formation between the two galaxies. In \citet{Monteagudo2018lmcsfh}, SFHs for several small fields in the central parts of the LMC were released, but again contained no metallicity information. \cite{ruiz-lara2020} used the deep, multi-band, contiguous SMASH data \citep{nidever2017smash} in the LMC to derive spatially-resolved SFHs. The paper was focused on the formation of the LMC's spiral arm and also did not include any metallicity information. Finally, \cite{Mazzi2021vmclmcsfh} used infrared VMC data \cite{Cioni2011VMC} covering the entire main body of the LMC. Unfortunately, infrared photometry is even less well suited than optical photometry for metallicity determination because it is less sensitive to metallicity, and the study does not include any AMR results. But they state that populations of stars made up of younger ages should have a lower metallicity than what is predicted from \cite{carrera2008lmcchem}. This is not to say that infrared photometry should not be used in studies of metallicity. In \cite{choudhury2021vmc} it was shown that metallicities derived from NIR photometry can be useful in exploring variations across the LMC. Therefore, establishing a precise and reliable source for an LMC AMR will benefit the current literature. Our goal is to take advantage of the precision in our metallicities and the extent of our data to generate an AMR that will cover more areas of the LMC to an unprecedented precision.

\begin{figure}
    \centering
    \includegraphics[width=0.45\textwidth]{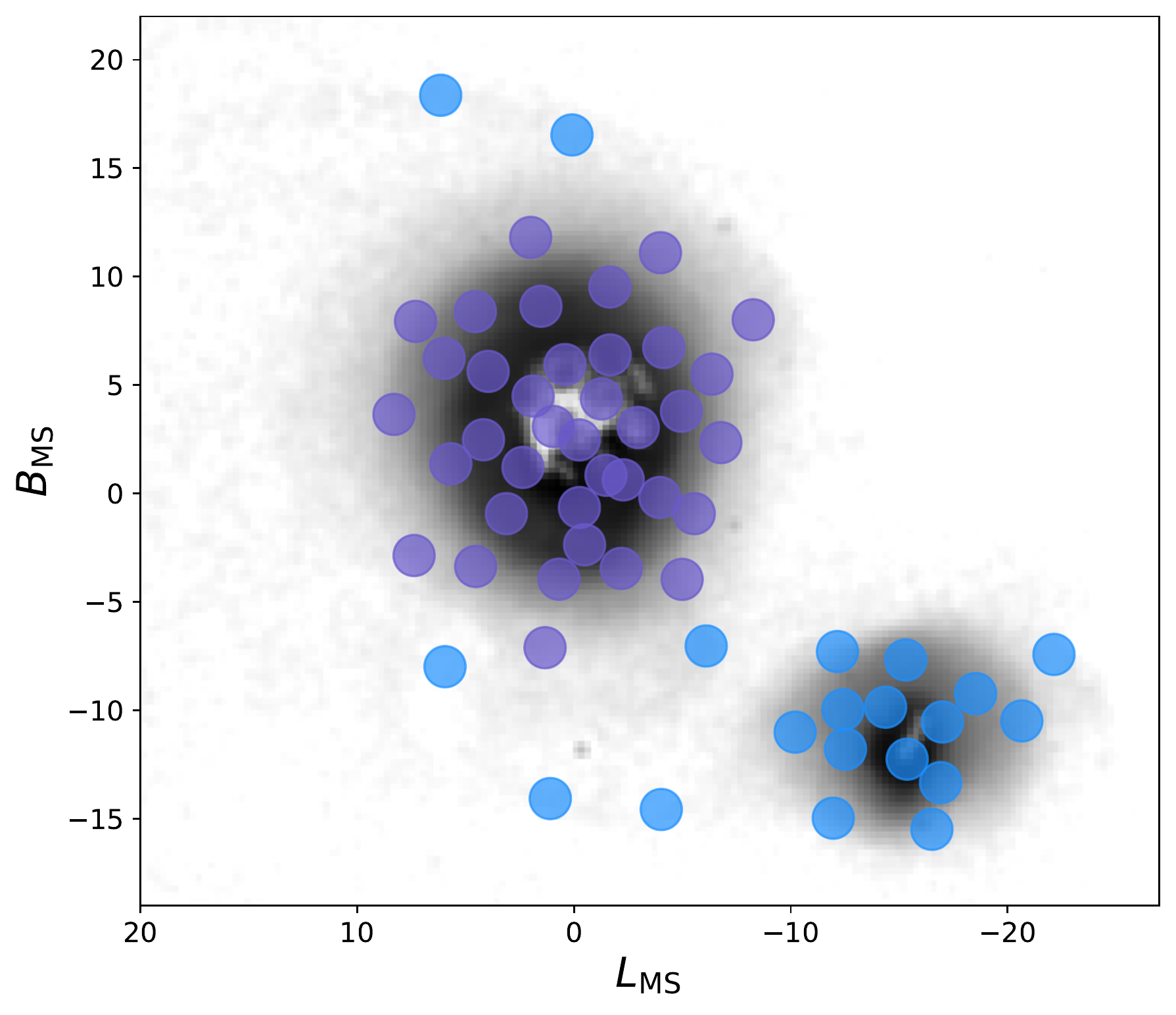}
    \caption{Map of the APOGEE-2 Magellanic Cloud fields filled circles) and the background showing the density of MC RGB stars selected with Gaia. The LMC fields analyzed here are shown as purple filled circles. For some basic statistics on the LMC fields see Table \ref{tab:ages_fields}.}
    \label{fig:ages_map}
\end{figure}

Alternatively, there have been many different studies of the LMC that specifically target the calculation of the AMR in the past. For example \cite{olszewski1991spectroscopy,dirsch2000age,grocholski2006ii} derive the AMR using clusters, while some papers such as \cite{carrera2008lmcchem} derive the AMR using spectroscopy of field stars. Most previous studies have a preference for clusters instead of individual stars and so with the derived individual ages and the known metallicities, we also explore the AMR of the LMC. 

One drawback with using clusters for AMR and other age studies is that an age gap exists creating a bimodal cluster age distribution \citep{dacosta1991clusters,geisler1997oldclusters}. Only two clusters in the LMC have been confirmed having ages between $\sim$3$-$12 Gyr old. These being ESO 121-SC 03 \citep{mateo1986ccd,olszewski1991spectroscopy} and KMHK 1592 \citep{piatti2022genuine} respectively. Having only two clusters for such a wide range of ages leads to large uncertainties in the behavior of the LMC for intermediate ages. However, \cite{gatto2020agegap} make the claim using YMCA (Gatto et al. in prep) and STEP \cite{ripepi2014step} data that the gap is observational bias. That work finds 16 candidate intermediate age clusters, which if confirmed will change our understanding of how star clusters formed in the LMC. Currently the only way to get data for intermediate ages is to use field stars to fill in the age gap where clusters are not present or to confirm the existence of more so called age gap clusters. 






This paper starts in Section \ref{sec:ages_data} by presenting the four datasets used. Sections \ref{sec:ages_lmcdist}, \ref{sec:ages_extlaw}, and \ref{sec:ages_salaris} introduce how distances to the individual LMC stars are calculated, the extinction laws used, and the Salaris correction to the metallicity, respectively. Next, Section \ref{sec:ages_extagemass} outlines how the extinctions, ages, and masses are calculated for each star with the validation explained in Section \ref{sec:ages_kasccalval}. In Section \ref{sec:ages_biascorr}, the bias correction applied to the LMC stars is discussed. Finally, the LMC results are presented in section \ref{sec:ages_lmcresults} and discussed in section \ref{sec:ages_discconc}. 

\section{Data} 
\label{sec:ages_data}

\subsection{APOGEE} 
\label{ssec:ages_apogeedata}

The Apache Point Observatory Galactic Evolution Experiment (APOGEE, \citealt{majewski2017apache}) is a valuable tool to study the LMC. APOGEE is part of the broader Sloan Digital Sky Survey (SDSS IV, \citealt{blanton2017sloan}).
APOGEE's main goal is to accurately and extensively study the chemistry and kinematics of the Milky Way (MW). It does this with two identical $H$-band spectrographs \citep{wilson19spectro} in the northern and southern hemispheres. For the northern hemisphere the spectrograph (APOGEE-N) is located at the Apache Point Observatory (APO) and takes data on the Sloan 2.5 m \citep{gunn2006apo} and NMSU 1.0 m \citep{holtzman101m} telescopes. As for the southern hemisphere, the spectrograph (APOGEE-S) is located at the Las Campanas Observatory (LCO) and connected to the du Pont 2.5 m telescope \citep{bowen1973optical}.

The raw APOGEE are processed with the data processing pipeline \citep{nidever15pipe} producing 1-D extracted and wavelength calibrated spectra with accurate radial velocities (RVs) determined with \texttt{Doppler} \citep{doppler}. Spectral parameters and abundances are then determined with the APOGEE Stellar Parameters and Chemical Abundances Pipeline (ASPCAP, \citealt{garciaperez16aspcap}).
This compares the normalized observed spectra with a large grid of synthetic spectra using \texttt{FERRE} \citep{allendeprieto2006}. The most up-to-date version of the data used in this work is part of SDSS-IV Data Release 17 \citep{Abdurro'uf2022ApJSdr17}. In particular for DR17, the APOGEE spectral grid was created using \texttt{synspec} \citep{hubeny21synspec}. Updates specific to DR17 can be found in Holtzmann et al. (in prep).

The main ASPCAP spectral fitting determines the ``spectral parameters'' which affect a significant fraction of the spectrum: \teffe, \logge, v$_{\text{micro}}$, [M/H], [C/M], [N/M], [$\alpha$/M], and v$_{\text{macro}}$ for giant stars. Individual elemental abundances are then determined from the same spectral grid by holding the spectral parameters constant but varying [M/H] (or [$\alpha$/M], depending on the element) with the use of narrow spectral windows specific to each element.
Chemical abundances provided in DR17 from ASPCAP are: C, C\textsc{i}, N, O, Na, Mg, Al, Si, S, K, Ca, Ti, Ti\textsc{ii}, V, Cr, Mn, Fe, Co, Ni, and Ce. For this work we extensively use \teffe, [Fe/H], and [$\alpha$/Fe] (=[$\alpha$/M]+[M/H]-[Fe/H]) derived by ASPCAP. Uncalibrated \teffe, \logge, vmicro, [M/H], [C/M], [N/M], [$\alpha$/M], and v $\sin$i can be found in each star's \texttt{FPARAM} array in the data catalog.

There are both systematic and statistical errors present in the APOGEE data. To minimize the effect of systematic errors, stars with known solar-like metallicity in the solar neighborhood were observed. From these stars offsets were derived and corrected for in the APOGEE data. Statistical uncertainties were calculated by performing multiple visits of single stars and fitting function of \teffe, [M/H], and SNR to the observed scatter. More information on $\alpha$-abundance uncertainties can be found in \cite{nidever20lazy}.

The LMC selection for DR17 is almost identical to that described in \citet{nidever20lazy} for DR16. The biggest difference is the coverage that has been extended to roughly twice the previous size (compare Fig.\ \ref{fig:ages_map} here to Fig.\ 1 in \citealt{nidever20lazy}). The quality cuts applied to each field star were:

\begin{itemize}
    \item \texttt{Teff$\_$BAD}, \texttt{LOGG$\_$BAD}, \texttt{VMICRO$\_$BAD}, \texttt{ALPHAFE$\_$BAD}, \texttt{CFE$\_$BAD}, \texttt{NFE$\_$BAD}, \texttt{NO$\_$ASPCAP$\_$RESULT} flags are not set
    \item \teff $< 5200\,$K
    \item \logg $< 3.4$
    \item S/N $> 20$.
\end{itemize}

\begin{figure*}
    \centering
    \includegraphics[width=\textwidth]{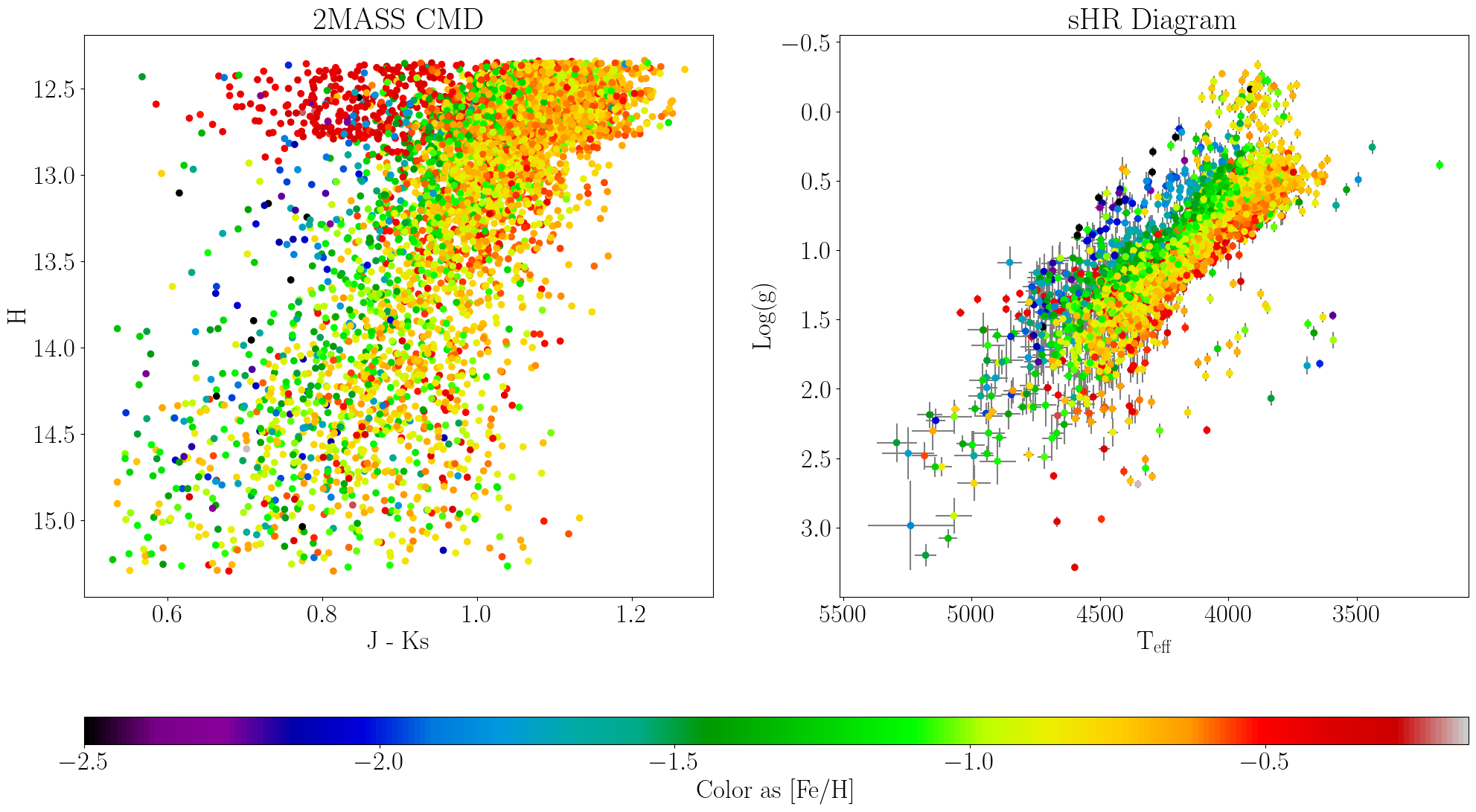}
    \caption{({\em Left}) A 2MASS color magnitude diagram of the selected APOGEE LMC RGB stars colored by [Fe/H]. The tip of the RGB is assumed to be at $H \sim$12.35 for the LMC \protect\citep{nidever20lazy}. ({\em Right}) A Kiel diagram for the APOGEE LMC RGB stars colored by [Fe/H].}
    \label{fig:ages_lmchr}
\end{figure*}

 After the membership selection and applying the above quality cuts, the total number of stars in the LMC sample is 6130. A CMD and HR diagram for the LMC RGB star sample can be seen in Figure \ref{fig:ages_lmchr}. 

Detailed information on the APOGEE-1/-2 targeting can be found in \cite{zasowski13target} and \cite{zasowski17target}. APOGEE DR17 has excellent spatial coverage of a significant portion of the LMC as seen in Figure \ref{fig:ages_map} out to a radius $\sim$10\dgr (or $\sim$8 kpc). More information on the fields can be seen in Table \ref{tab:ages_fields}.

\begin{table*}
    \label{tab:ages_fields}
    \caption{A table of positional data for each of the APOGEE LMC fields including the number of RGB stars observed. R and PA are the projected radius and position angle respectively. N$_{\rm Brt}$ is the number of identified bright RGB stars while N$_{\rm Fnt}$ is the number of faint sources (see Section~\ref{sec:ages_biascorr}).}
    \begin{tabular}{cccccccc}
        \hline
        Name & R.A. & Decl. & R & PA & N$_{\rm RGB}$ & N$_{\rm Brt}$ & N$_{\rm Fnt}$\\
         & (J2000.0) & (J2000.0) & (deg) & (deg) & & & \\
        \hline
        30Dor & 05:34:3.30 & -69:21:6.40 & 0.8& 52.0 & 41 & 41 & 0 \\
        LMC1 & 04:14:32.0 & -71:59:22.2 & 6.3 & 242.0 & 169 & 10 & 159 \\
        LMC2 & 04:13:14.9 & -68:28:21.6 & 6.7 & 272.0 & 148 & 9 & 139 \\
        LMC3 & 04:49:10.2 & -75:11:48.9 & 6.1 & 204.0 & 188 & 26 & 162 \\
        LMC4 & 04:54:54.6 & -68:48:6.80 & 3.1 & 287.0 & 254 & 254 & 0 \\
        LMC5 & 04:57:10.3 & -71:08:54.2 & 2.9 & 240.0 & 236 & 236 & 0 \\
        LMC6 & 05:10:55.5 & -65:46:21.7 & 4.4 & 337.0 & 200 & 200 & 0 \\
        LMC7 & 05:14:46.4 & -62:43:22.3 & 7.3 & 349.0 & 171 & 62 & 109 \\
        LMC8 & 05:20:48.5 & -72:32:48.2 & 2.8 & 191.0 & 220 & 220 & 0 \\
        LMC9 & 05:22:15.7 & -69:48:1.30 & 0.7 & 260.0 & 143 & 143 & 0 \\
        LMC10 & 05:30:28.3 & -75:53:11.2 & 6.1 & 178.0 & 171 & 35 & 136 \\
        LMC11 & 05:41:49.0 & -63:37:17.1 & 6.4 & 14.0 & 195 & 128 & 67\\
        LMC12 & 05:44:9.90 & -60:38:40.1 & 9.4 & 13.0 & 138 & 17 & 121 \\
        LMC13 & 05:44:24.1 & -67:41:59.7 & 2.7 & 38.0 & 173 & 173 & 0 \\
        LMC14 & 05:50:34.4 & -70:58:29.1 & 2.2 & 122.0 & 152 & 152 & 0 \\
        LMC15 & 06:08:2.80 & -63:47:24.5 & 7.3 & 38.0 & 175 & 47 & 128 \\
        LMC16 & 06:29:7.50 & -75:04:58.8 & 6.9 & 145.0 & 131 & 13 & 118 \\
        LMC17 & 06:29:59.5 & -70:17:23.1 & 5.3 & 102.0 & 176 & 44 & 132 \\
        LMC18 & 06:06:28.4 & -66:26:29.5 & 5.0 & 51.0 & 192 & 192 & 0 \\
        LMC19 & 06:12:35.8 & -69:27:47.4 & 4.0 & 90.0 & 182 & 182 & 0 \\
        LMC20 & 05:14:31.3 & -74:25:36.5 & 4.7 & 191.0 & 208 & 208 & 0 \\
        LMC21 & 04:35:3.80 & -67:39:8.40 & 5.3 & 289.0 & 205 & 118 & 87 \\
        LMC22 & 04:35:47.4 & -71:37:2.20 & 4.7 & 242.0 & 205 & 167 & 38 \\
        LMC23 & 07:07:56.0 & -73:45:55.5 & 8.6 & 128.0 & 86 & 4 & 82 \\
        LMC24 & 05:44:24.8 & -79:06:12.0 & 9.3 & 175.0 & 70 & 4 & 66 \\
        LMC25 & 06:33:14.1 & -63:39:58.6 & 8.9 & 54.0 & 170 & 15 & 155 \\
        LMC26 & 04:15:7.10 & -62:35:31.7 & 10.1 & 307.0 & 65 & 4 & 61 \\
        LMC27 & 05:12:55.0 & -67:59:33.1 & 2.3 & 321.0 & 152 & 152 & 0 \\
        LMC28 & 04:56:36.5 & -60:48:53.1 & 9.6 & 337.0 & 185 & 11 & 174 \\
        LMC29 & 06:23:21.8 & -65:40:18.9 & 6.7 & 58.0 & 210 & 79 & 131 \\
        LMC30 & 04:50:5.40 & -65:11:22.6 & 5.9 & 318.0 & 205 & 125 & 80 \\
        LMC31 & 05:31:58.7 & -66:27:10.7 & 3.5 & 13.0 & 152 & 152 & 0 \\
        LMC32 & 04:27:18.9 & -65:41:49.5 & 7.1 & 299.0 & 221 & 24 & 197 \\
        LMC33 & 06:51:56.5 & -67:24:9.50 & 8.1 & 83.0 & 180 & 12 & 168 \\
        LMC34 & 06:05:43.3 & -72:55:21.2 & 4.4 & 140.0 & 201 & 201 & 0 \\
        LMC35 & 04:06:26.5 & -74:54:2.80 & 7.9 & 221.0 & 160 & 7 & 153 \\
        \hline
    \end{tabular}
\end{table*}

\subsection{Gaia EDR3}
\label{ssec:ages_gaiadata}

The Gaia mission \citep{gaia2016gaia,gaia2021gaia} is an extensive space-based all-sky survey. We use both photometric and astrometric data from the early data release 3 (Gaia EDR3), which contains over 1.8 billion sources.  Gaia obtains photometric data in three optical bands: denoted ${BP}$, $G$, and ${RP}$. All of the Gaia EDR3 data are publicly available online from the Gaia Archive\footnote{\url{https://gea.esac.esa.int/archive/}}. The APOGEE DR17 catalog includes the corresponding Gaia photometry and astrometry columns and much of Gaia EDR3 and Gaia DR3 \citep{gaia2022dr3} are the same for the LMC stars in this work. 

\begin{figure}
    \centering
    \includegraphics[width=0.45\textwidth]{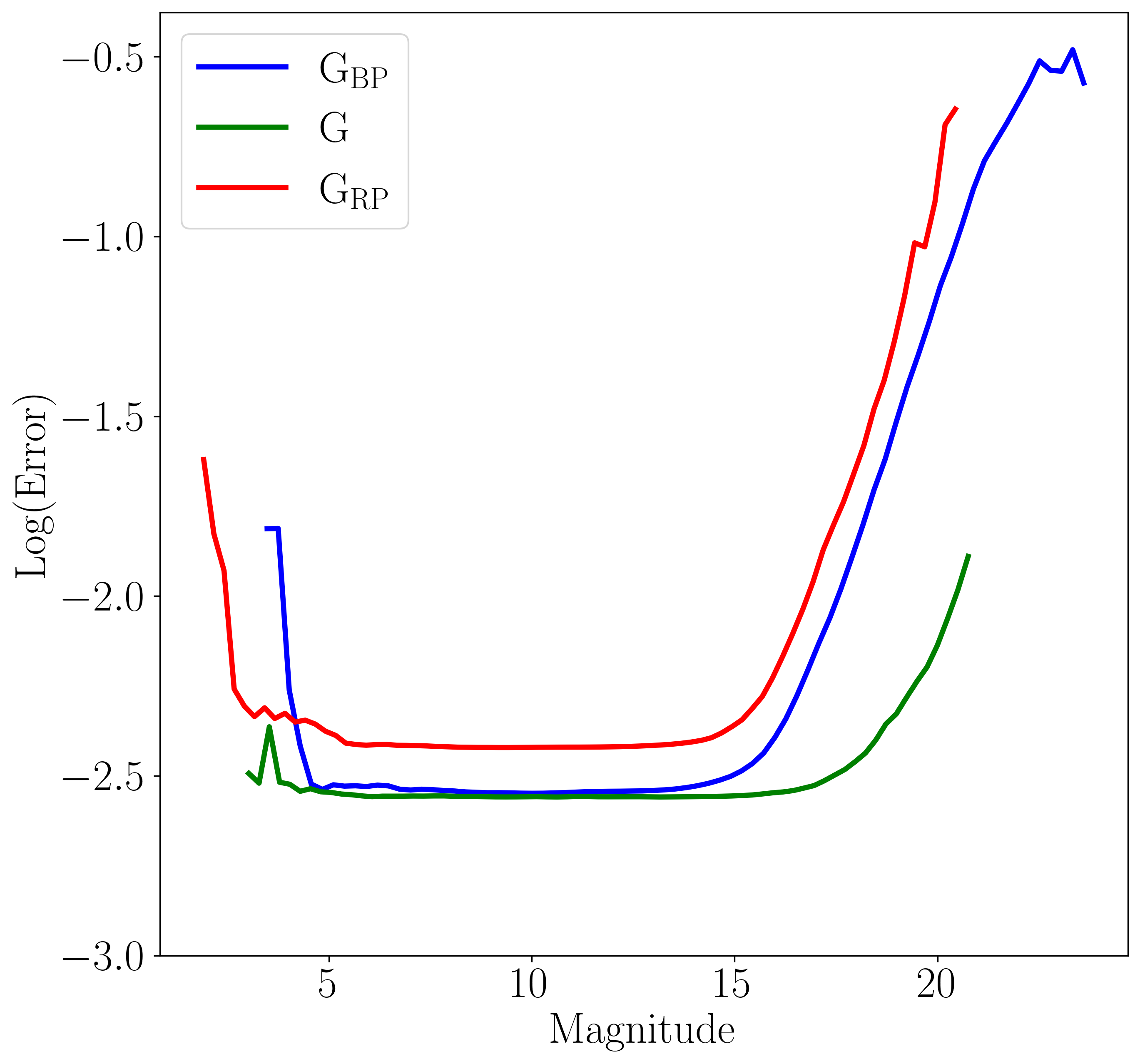}
    \caption{A reproduction of the magnitude error curve for Gaia EDR3 calculated using the Gaia photometry and equation \ref{equ:magerr}. The blue, green, and red curves represent the BP, G, and RP bands respectively. The curves here match qualitatively to the curves in \protect\cite{riello2021phot}.}
    \label{fig:ages_gaiaerror}
\end{figure}

Our age calculations in section \ref{ssec:ages_agemethod} require uncertainties in the photometric magnitudes.  We calculate these from the relative flux uncertainties provided in the data release catalog.
Flux and magnitude are related by $m_\lambda = -2.5\log(F_\lambda)$, where $m_\lambda$ is the magnitude and $F_\lambda$ is the flux. Using general error propagation methods and taking into account the zero point for each Gaia band, the magnitude error is given by

\begin{equation}\label{equ:magerr}
    \sigma_{m\lambda} = \sqrt{\bigg(\frac{1.085}{F_{\lambda}/\sigma_{\lambda}}\bigg)^2+zp_\lambda^2}
\end{equation}

\noindent
where $\sigma_{m\lambda}$ is the magnitude uncertainty for the $\lambda$ Gaia band, $F_{\lambda}/\sigma_{\lambda}$ is the flux over its error provided by Gaia, and $zp_\lambda$\footnote{See \url{https://www.cosmos.esa.int/web/gaia/dr3-passbands} for these values} is the zero point offset for the band. Plotting the calculated error as a function of the magnitude produces the curve in Figure \ref{fig:ages_gaiaerror}. This was created by taking a large selection of stars and interpolating the error as a function of magnitude. Qualitatively this figure closely resembles the analogous figure in \cite{riello2021phot}.

Gaia EDR3 includes very accurate proper motions ($\mu$) and parallaxes ($\varpi$). The reciprocal of the Gaia parallax is used to calculate distances of stars in the APOKASC validation set (see section \ref{ssec:ages_apokascdata}).  While taking the parallax reciprocal is non-optimal for low-S/N values, the large majority of stars in the APOKASC sample have S/N$\gtrsim$70 where this is less of a problem.


\subsection{APOKASC}
\label{ssec:ages_apokascdata}

APOKASC is a joint spectroscopic and asteroseismic dataset created by combining APOGEE and \textit{Kepler} data \citep{pinsonneault2014apokasc,pinsonneault2018apokasc}.  The APOKASC stars have well-determined masses, radii and ages.  We use the APOKASC 3 (Pinsonneault et al. in prep) catalog to validate our estimated ages and masses.

\begin{figure}
    \centering
    \includegraphics[width=0.475\textwidth]{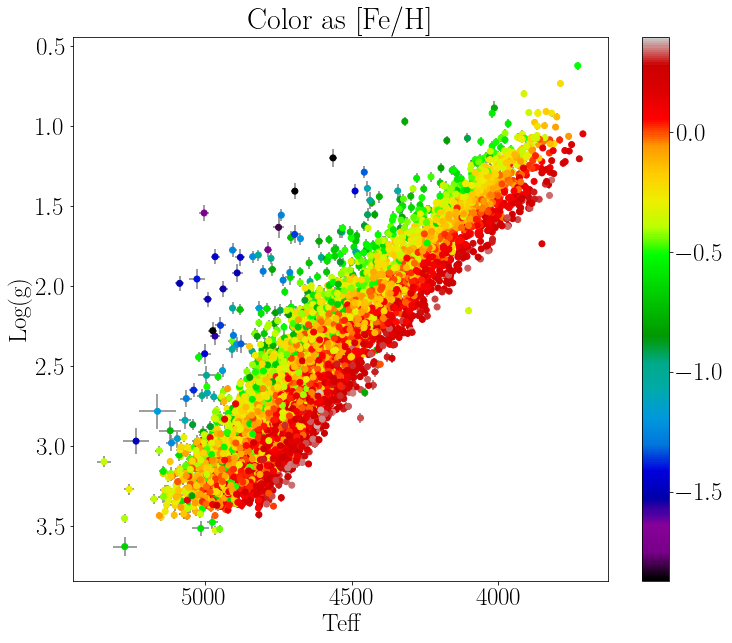}
    \caption{An HR diagram for the selected APOKASC RGB stars colored by [Fe/H].}
    \label{fig:ages_kaschr}
\end{figure}

The following selection criteria are applied to the APOKASC data to best match the APOGEE data for the LMC sample:
\begin{itemize}
    \item \texttt{APOKASC3\_CONS\_EVSTATES} = 1
    \item $\varpi/\sigma_\varpi > 3.0$
\end{itemize}

\noindent
where \texttt{APOKASC3\_CONS\_EVSTATES} denotes RGB stars in the 6.7.4 version of the APOKASC dataset (in contrast to red clump stars). The parallax cut is performed to ensure accurate distances can be found by taking the reciprocal. After the selection criteria is applied, there are 4058 RGB stars left in the APOKASC sample that span \teff from 3712 to 5345. A Kiel diagram of the final selection is shown in Figure \ref{fig:ages_kaschr}.


\subsection{PARSEC Isochrones}
\label{ssec:ages_parsecdata}


In order to derive extinction, age, and mass estimates of stars we rely on the use of PARSEC (\citealt{bressan2012parsec}; \citealt{marigo2017}) isochrones, which are available online\footnote{\url{http://stev.oapd.inaf.it/cgi-bin/cmd}}. These contain \teffe, \logge, metallicity, initial mass, the photometric absolute magnitudes, as well as a convenient stellar evolutionary phase label ({\tt LABEL}).  We downloaded a finely-spaced grid of isochrones in age and metallicity for the Gaia EDR3 and 2MASS bands.  The ages span 25 Myr to 17 Gyr in steps of 50 Myr, and the metallicities span [M/H]=$-$2.3 to $+$0.5 in steps of 0.02 dex.


\section{LMC Distances} 
\label{sec:ages_lmcdist}
The geometry of the LMC disk is well-modeled as a thin, inclined plane. Using this simple model and well-constrained observational parameters, the distances for the LMC disk stars can be calculated. The mathematical basis of this model comes from \cite{van2001magellanic} and \cite{choi2018smashing}.

To calculate the distance in the inclined disk model, the first step is to convert the ($\alpha$, $\delta$) celestial coordinates to a cylindrical coordinate system ($\rho$, $\phi$). These coordinates are defined such that $\rho$ is the angular distance from the LMC center (radius) and $\phi$ is the position angle (east of north). This is analogous to polar coordinates in 2D, but on the celestial sphere. The coordinate conversion to the cylindrical coordinates is:
\begin{equation} \label{equ:angcoor}
    \begin{array}{l}
    \cos{\rho} = \cos{\delta_0}\cos{\delta}\cos{(\alpha-\alpha_0)} + \sin{\delta_0}\sin{\delta}, \\
    \sin{\rho}\cos{\phi} = -\cos{\delta}\sin{(\alpha-\alpha_0)}, \\
    \sin{\rho}\sin{\phi} = \cos{\delta_0}\sin{\delta} - \sin{\delta_0}\cos{\delta}\cos{(\alpha-\alpha_0)},
    \end{array}
\end{equation}
where $(\alpha_0, \delta_0) = (82.25^\circ, -69.50^\circ)$ is from \cite{van2001magellanic}.

The distance can be calculated by projecting the angular coordinates onto a plane which means the distance is the given by

\begin{equation}\label{equ:dist}
    D = \frac{D_0\cos{i}}{\cos{i}\cos{\rho} - \sin{i}\sin{\rho}\sin{(\phi-\theta)}},
\end{equation}

\noindent
where $D_0 = 49.9$ kpc \citep{van2001magellanic, degrijs2014clustering} is the distance from the Sun to the center of the LMC, $i = 25.87^\circ$ is the inclination of the LMC disk, and $\theta = 149.23^\circ$ is the position angle of the line of nodes. Both the inclination and position angle of the line of nodes comes from the \cite{choi2018smashing} analysis of the SMASH \citep{Nidever2021} red clump stars. When using equation \ref{equ:dist}, it is important to note that many angle are measured from North and not East, so $90^\circ$ is added to $\theta$ in practice.

\begin{figure}
    \centering
    \includegraphics[width=0.45\textwidth]{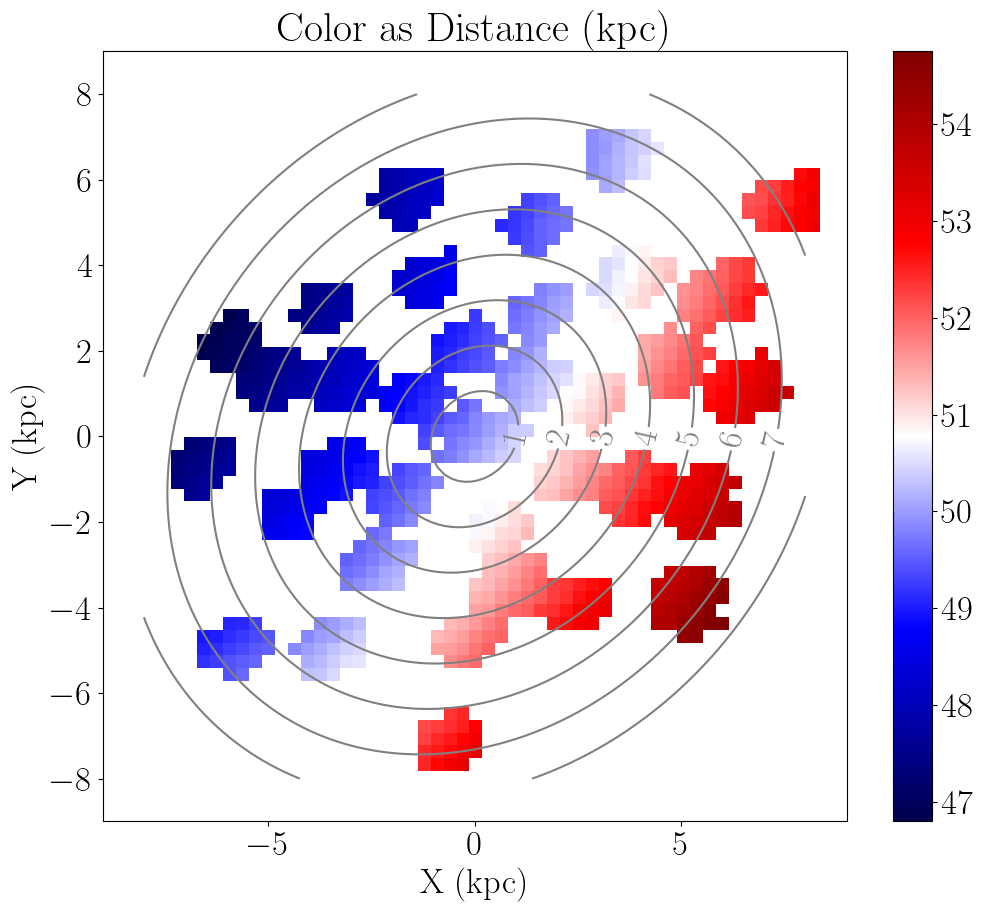}
    \caption{The resulting distance map for the LMC using Equation \ref{equ:dist}. The contours show lines of constant ``elliptical'' radius starting at 1 kpc for the center contour out to 8 kpc calculated using equation 6 from \protect\cite{choi2018smashing}}.
    \label{fig:ages_lmcdist}
\end{figure}

We take the equation for finding the radius directly from \cite{choi2018smashing}, which is
\begin{equation}\label{equ:ellrad}
    r(x,y)^2 = (x\cos{\psi} - y\sin{\psi})^2 + \bigg(\frac{x\sin{\psi} + y\cos{\psi}}{b/a}\bigg)^2,
\end{equation}
\noindent
where $\psi\,(= 227.24^\circ)$ is the position angle of the semi-major axis, and $b/a\,(= 0.836)$ is ratio of the semi-minor axis to the semi-major axis. For angles measured from the north, $90^\circ$ must be added to $\psi$ just like with $\theta$ above. A distance map of the LMC with the added elliptical radius contours (annuli) can be seen in Figure \ref{fig:ages_lmcdist}.

\section{Extinction Laws} 
\label{sec:ages_extlaw}

\subsection{Cardelli Extinction Law}
\label{ssec:ages_ccm89}

In order to calculate and constrain the extinction across multiple bands, we adopt the extinction law directly from \citet[][hereafter CCM89]{cardelli89extinction} for calculating the extinction coefficients. 

The extinction relative to the $V$ band can be found with
\begin{equation}
\label{equ:relext}
    \text{A}_\lambda/\text{A}_V = a_\text{CCM}(x) + b_\text{CCM}(x)/R_V,
\end{equation}
\noindent
 where $a_\text{CCM}(x)$ and $b_\text{CCM}(x)$ are functions of wavelength ($x = 1/\lambda$) with functional forms that vary with the particular wavelength regime (i.e. IR, Optical) and $R_V \equiv \text{A}_V/\text{E}(B-V)$.

For the IR regime ($0.3 \leq x \leq 1.1\, \mu m^{-1}$), $a_\text{CCM}(x)$ and $b_\text{CCM}(x)$ are simply
\begin{subequations}
\label{equ:relext_ir}
    \begin{align}
        a_\text{CCM}(x) &= 0.574x^{1.61}\\
        b_\text{CCM}(x) &= -0.527x^{1.61}.
    \end{align}
\end{subequations}

For the NIR and Optical regime ($1.1 \leq x \leq 3.3\, \mu m^{-1}$), $a_\text{CCM}(x)$ and $b_\text{CCM}(x)$ are
\begin{subequations}
\label{equ:relext_niroptical}
\begin{align}
y(x) &= x - 1.82 \\
  \begin{split}
        a_\text{CCM}(y) &= 1 + 0.17699y - 0.50477y^2 - 0.02427y^3 \\
    & + 0.72085y^4 + 0.01979y^5 - 0.77530y^6 \\ 
    & + 0.32999y^7
  \end{split}
  \\
  \begin{split}
        b_\text{CCM}(y) &= 1.41338y + 2.228305y^2 +1.07233y^3 \\
    & - 5.38434y^4 - 0.62251y^5 +5.30280y^6 \\ 
    & - 2.09002y^7
  \end{split}
\end{align}
\end{subequations}

As noted in CCM89, a value of 3.1 for $R_V$ reproduces what would be expected for the diffuse interstellar medium and only starts to substantially deviate from this value in the UV regime ($\lambda <$ 0.303 $\mu$ m). Since we are only considering the optical and IR regimes, $R_V$ = 3.1 is used throughout this work unless, otherwise specified.


In this work, the $G$ band extinction (A$_G$) is taken as the fiducial instead of A$_V$ and this is accomplished by simply dividing Equation \ref{equ:relext} by A$_G$/A$_V$. A figure of extinctions of all six photometric bands relative to A$_G$ can be seen in Figure \ref{fig:ages_ccm89fitz99}. The CCM89 extinction law is included in the \texttt{dust$\_$extinction}\footnote{\url{https://dust-extinction.readthedocs.io/en/stable/\#}} Python package, which is used throughout this work for the actual calculation of the extinction coefficients.

\subsection{Fitzpatrick Extinction Law}
\label{ssec:ages_f99}

\begin{figure}
    \centering
    \includegraphics[width=0.45\textwidth]{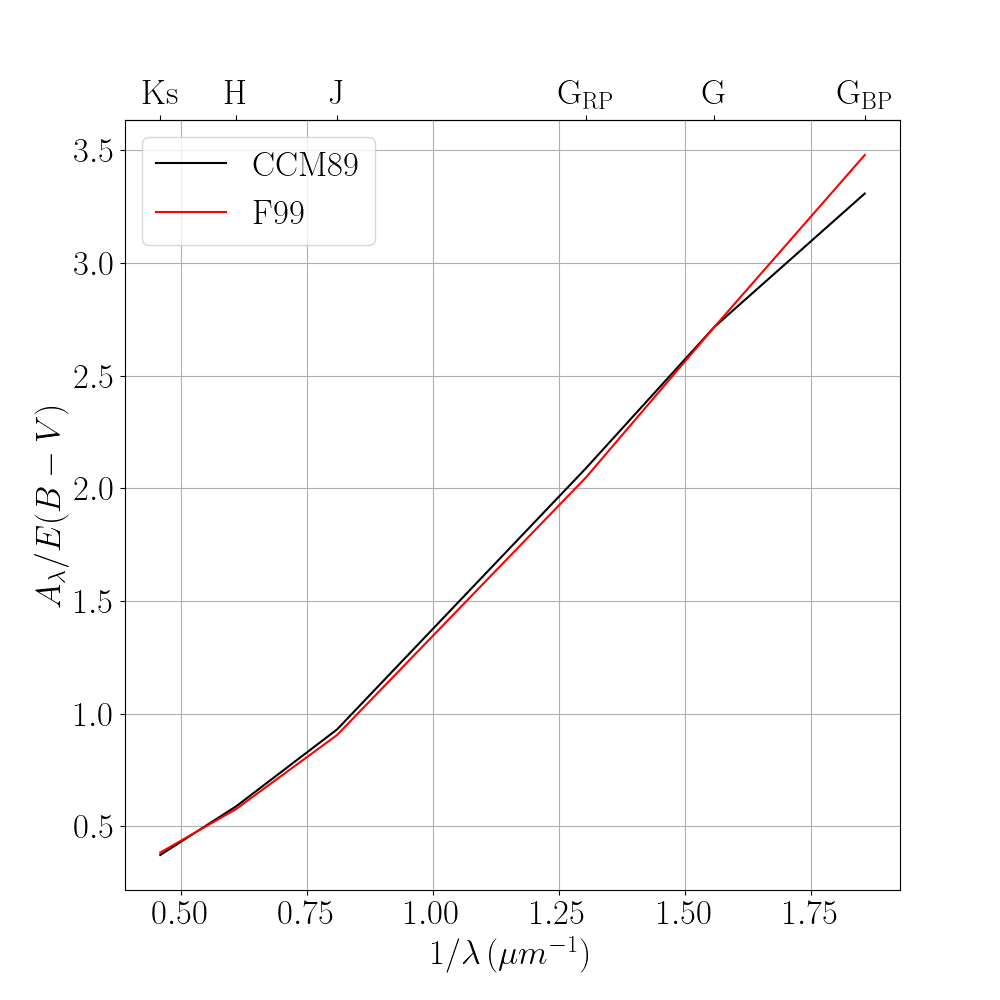}
    \caption{The relative extinction curve of $A_\lambda/E(B-V)$ from both CM89 and F99. The top axis shows the different photometric filters used in this work.}
    \label{fig:ages_ccm89fitz99}
\end{figure}

We make use of a second extinction law to validate the extinctions from \citet[][hereafter F99]{fitzpatrick1999extinction}. For a more detailed discussion on why F99 is used see Section \ref{ssec:ages_kascext}. Also, as with CCM89, we make use of \texttt{dust$\_$extinction} to calculate the extinction coefficients. While similar to CCM89, F99 gives slightly different results (see Figure \ref{fig:ages_ccm89fitz99}).

\section{Salaris Correction}
\label{sec:ages_salaris}

Most stellar isochrone models are based on a scaled solar composition, but in actuality stars have a variety of chemical compositions. One solution that allows the use of the solar scaled isochrones for stars with a more complex composition is to shift the [Fe/H] of the isochrone based on the star's [$\alpha$/Fe] abundance. Stars that are enhanced in $\alpha$-elements will appear to have a cooler \teff than expected. This can result in an older age being assigned to the star. \citet{salaris1993alpha} determined that an $\alpha$-abundance correction of the following form would fix this:

\begin{equation}
\label{equ:salaris}
    \text{[Fe/H]}_\text{sal} = \text{[Fe/H]} + \log(a_\text{sal} f_\alpha + b_\text{sal}),
\end{equation}

\noindent
where [Fe/H] is the uncorrected metallicity, $a_\text{sal}$ and $b_\text{sal}$ are coefficients and $f_\alpha = 10^{[\alpha/\text{Fe}]}$. The original values for $a_\text{sal}$ and $b_\text{sal}$ are 0.638 and 0.362, respectively \citep{salaris1993alpha}.

For this work we recalculate the $a_\text{sal}$ and $b_\text{sal}$ using the solar composition from \cite{asplund2021sun}. Calculating the coefficients is done using

\begin{subequations}
\label{equ:salariscoeff}
    \begin{align}
        a_\text{sal} &\equiv \sum_\text{X} \bigg(\frac{\text{X}}{\text{Z}}\bigg) = \sum_\text{X} 10^{\log\epsilon_X - 12.0}\frac{A_\text{X}}{A_\text{H}}\bigg(\frac{\text{Z}}{\text{X}}\bigg)_\odot^{-1} \\
        b_\text{sal} &= 1 - a_\text{sal}
    \end{align}
\end{subequations}

\noindent 
where $\log\epsilon_X$ is the value reported in \cite{asplund2021sun}, $A_X/A_H$ is the atomic mass ratio of element $X$ to hydrogen from IUPAC\footnote{\url{https://iupac.qmul.ac.uk/AtWt/}} (not to be confused with the relative extinction coefficients used elsewhere in this work), and $(Z/X)_{\odot}$ is the \cite{asplund2021sun} value of $0.0187 \pm 0.0009$. The sum is over the exact same elements used in \cite{salaris1993alpha} (i.e. O, Ne, Mg, Si, S, Ca). The new updated values for the correction are $a_\text{sal} = 0.659$ and $b_\text{sal} = 0.341$.

\begin{figure*}
    \centering
    \includegraphics[scale=0.3]{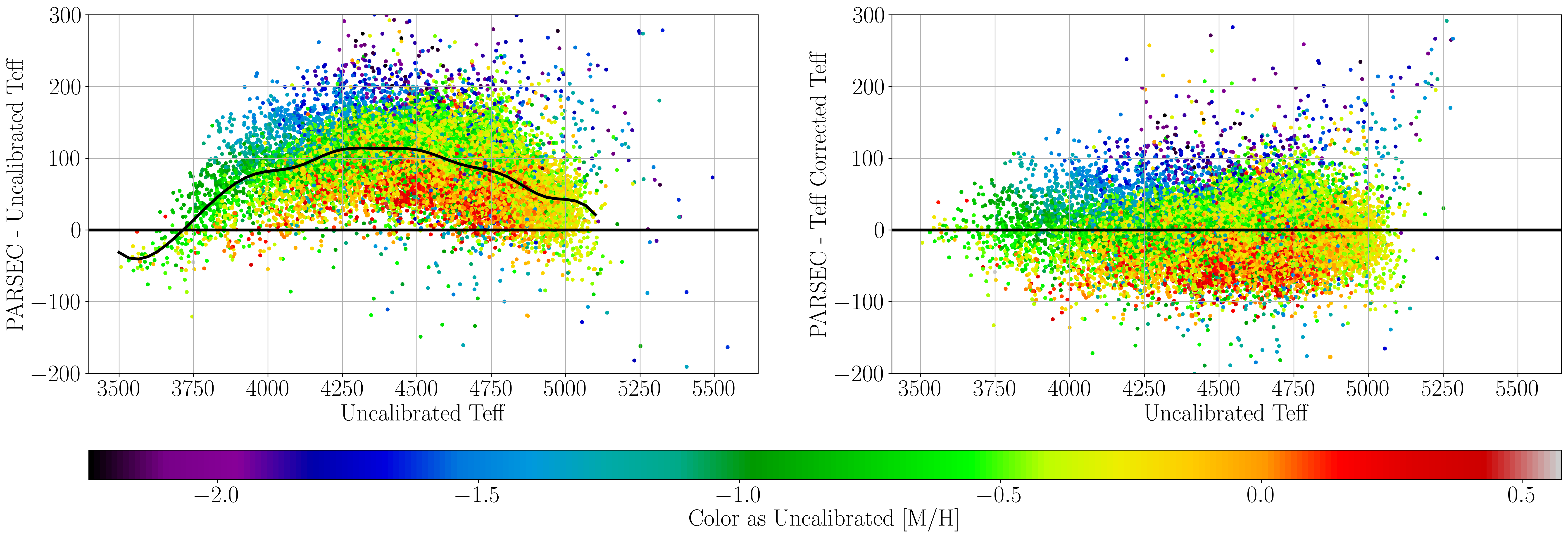}
    \caption{({\em Left}) The residuals of the photometric and uncalibrated ASPCAP \teff values with the spline used to temperature correct the uncalibrated \teffe. ({\em Right}) The residuals after the \teff correction has been applied. There is a clear trend in [M/H] and so a second spline as a function of [M/H] has been applied (see the text and Figure \ref{fig:ages_Teff_cal_mh}).}
    \label{fig:ages_Teff_cal_Teff}
\end{figure*}

\begin{figure*}
    \centering
    \includegraphics[scale=0.3]{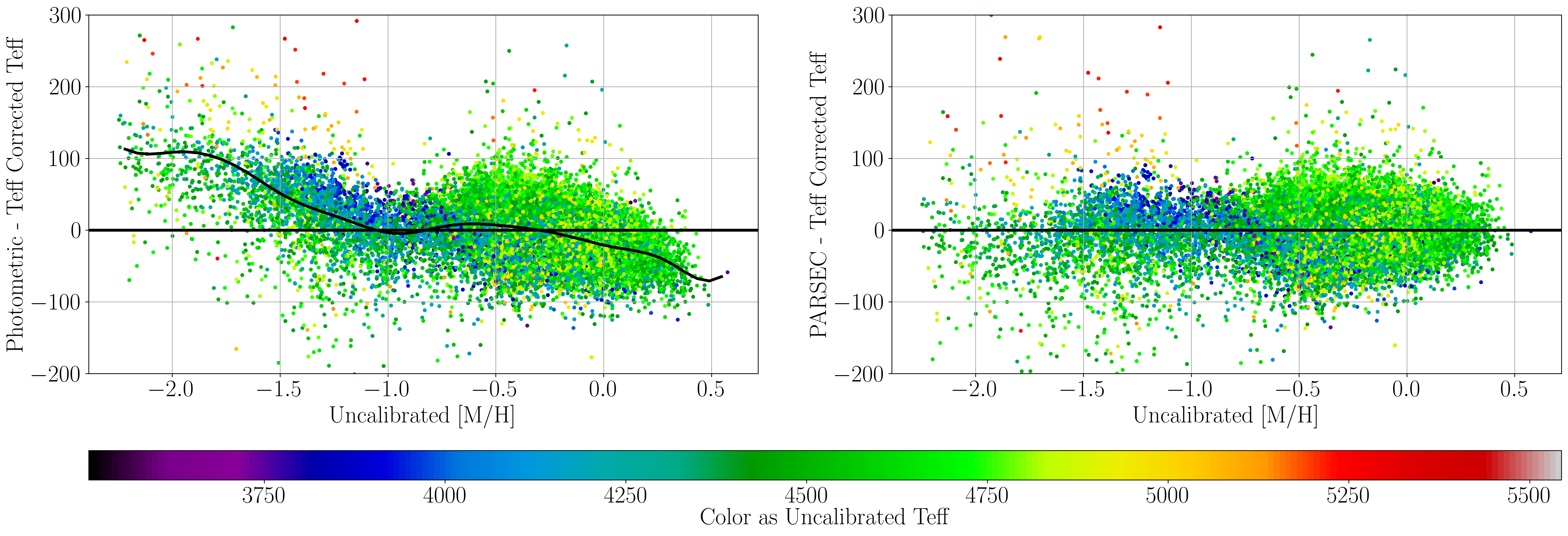}
    \caption{({\em Left}) A plot of the residuals between the photometric and 
    \teffe--corrected spectroscopic \teff values (\teff spline applied) with the spline used to [M/H] correct the \teff values. ({\em Right}) The residuals after the [M/H] correction has been applied. The median residual after the [M/H] has been applied is $-$1.72 K with a scatter of 23.75 K.}
    \label{fig:ages_Teff_cal_mh}
\end{figure*}

\section{Calibrations}

\subsection{Effective Temperature}
\label{ssec:ages_Teffcal}

Having an accurate temperature for a star is critical as the calculation of ages and extinctions in Section \ref{sec:ages_extagemass} relies heavily on this value. It is also important that the observed \teff and isochrone \teff values agree with each other.  We find that the results from our age validation sample is greatly improved when the temperatures provided by APOGEE are recalibrated and the ``named'' \texttt{TEFF} values are not used directly. 

Here we outline how the recalibration was performed with the stars in a two step process by using PARSEC isochrones and low extinction RGB stars in the full APOGEE \texttt{allStar} summary file\footnote{\url{https://www.sdss.org/dr17/irspec/spectro_data/}}. RGB stars were used for the calibration as the LMC stars should be RGB. First, the RGB stars were selected using the criteria


\begin{itemize}
    \item uncalibrated log(g) $<$ 3.5
    \item uncalibrated \teff $<$ 6000 K
\end{itemize}

\noindent
where the uncalibrated $\log(\text{g})$ and \teff values can be found in the \texttt{FPARAM} column in the \texttt{allStar} file. This selection does not remove red clump (RC) stars, which were identified and removed using 

\begin{itemize}
    \item 2.38 $<$ uncalibrated log(g) $<$ 3.5
    \item {[C/N]} $>$ 0.04 - 0.46[M/H] - 0.0028$\cdot dT$
\end{itemize}


\noindent
where 

\begin{equation}
    \begin{split}
        dT &= \text{Teff}_\text{spec} - (4400 - 552.6\cdot(\log \text{g}_\text{spec} - 2.5) \\
        & - 324.6\cdot\text{[M/H]})
  \end{split}
\end{equation}
\noindent
where $\text{T}_\text{eff,spec}$, and $\log \text{g}_\text{spec}$ are \texttt{Teff\_SPEC} and \texttt{LOGG\_SPEC} in the \texttt{allStar} file, respectively. These are the same RGB selection criteria from DR16 used for the \teff calibration \citep{jonsson2020data}. Any star with a non-zero \texttt{ASPCAPFLAG} value was removed to create a ``pristine" sample of RGB stars. The pristine sample overlaps the RGB stars in both the APOKASC and the selected LMC RGB datasets in \teff--\logg--[Fe/H].

Once the RGB stars were selected, low extinction stars were picked out such that \texttt{SFD\_EBV} $\leq$ 0.03, where \texttt{SFD\_EBV} is the $E(B - V)$ value from \cite{schlegel1998reddening}. In actuality, the value of the cutoff is unimportant so long as it is small so that the attenuation due to the reddening is negligible when calculating the intrinsic color for calibration. There are some regions where the Schlegel map does not give accurate values for reddening, such as the center of the LMC. It is assumed that the calibration RGB stars cover enough of the sky where the Schlegel map is accurate that stars in places where this is not true can still be calibrated with the method in this section.  

Even after making the necessary cuts to create a sample of RGB stars, there are still some RC stars remaining. These stars are easily removed because they do not follow the expected RGB trend in \logge--\teff--[Fe/H] space.  



Next, the photometric temperatures were calculated for the sample of low extinction RGB stars using the PARSEC isochrones through the following process:


\begin{enumerate}
    \item Pick all isochrones with the Salaris corrected metallicity of the star. 
    \item Remove extremely old (10 Gyr < age) or young ages (age > 0.5 Gyr). Age in itself does not greatly impact the color--\teff relation for the isochrones.
    \item The isochrone color--\teff relation is interpolated using a B-spline for each of the colors (i.e., ${BP} - G$, $G - {RP}$, $G - J$, $G - H$, and $G - K_{\rm s}$).
    \item The ``photometric \teffe'' for each star is then calculated using the weighted mean of the \teff values from the individual colors.
\end{enumerate}



Calibrating the \teff values involves fitting residuals between the photometric and spectroscopic temperatures. We find the best way to calibrate the \teff values is a two-step process. The first step is fitting a spline as a function of the uncalibrated spectroscopic \teff values to the median residuals (photometric \teff $-$ uncalibrated spectroscopic \teffe). In this case, the spline is fit to the binned median values. Figure \ref{fig:ages_Teff_cal_Teff} shows the \teff residuals before and after of this first  step. After this initial correction there is a trend in [M/H] that is still present and, therefore, the second step is to fit a second spline as a function of uncalibrated [M/H] to the median residuals of the photometric \teff and the \teffe--corrected spectroscopic \teffe. Figure \ref{fig:ages_Teff_cal_mh} illustrates this second step. In the end, the final photometric and calibrated spectroscopic \teff residuals have a median of $-$1.72 K and scatter of 23.75 K. The splines derived from fitting the residuals were then used to calculate the \teff values for the datasets on this work.

\subsection{APOKASC Ages}
\label{ssec:ages_kascagecal}

The ages presented in the APOKASC catalog were derived 
using isochrones from the Yale Rotating Evolution Code \citep[YREC,][]{pinsonneault1989evolutionary,van2012sensitivity} or the Garching Stellar Evolution Code \citep[GARSTEC,][]{weiss2008garstec} not the PARSEC isochrones as we use here.  For consistency's sake, we re-derive ages for the APOKASC stars using their APOKASC-determined mass, the star's Salaris corrected metallicity, and the PARSEC isochrones using the following procedure:

\begin{enumerate}
    \item  All isochrones with the two closest metallicity values to the star's own Salaris corrected metallicity are selected.
    \item For both the lower and higher metallicities, the age is interpolated as function of mass for the APOKASC-mass value of the star.
    \item The ages for the lower and higher metallicites are linearly interpolated to the star's Salaris corrected metallicity giving its final age.
\end{enumerate}





The calculation of the APOKASC ages using the PARSEC models and the APOKASC 3 masses versus the original APOKASC 3 ages do differ slightly. On average the re-derived APOKASC ages tend to be younger. The youngest ages remain relatively unchanged while for the oldest ages the deviation is $\sim$1.75 Gyr.

\section{Extinction, Age, \& Mass Calculation}
\label{sec:ages_extagemass}


\subsection{Extinction Calculation}
\label{ssec:ages_extmethod}

Extinction is the attenuation of the flux of a star due to dust and gas between the observer and the star. This needs to be accounted for in order to produce accurate absolute photometry as the age method described in Section \ref{ssec:ages_agemethod} depends sensitively on the photometry.

An important quantity directly related to the extinction is the reddening. The reddening of a stars is the color excess due to different amounts of extinction in two different bands.

\begin{equation}
E(G - K_{\rm s}) = (G - K_{\rm s})_\text{obs} - (G - K_{\rm s})_\text{int} = A_{\rm G} - A_{\rm Ks}
\end{equation}

\noindent
Using the $G$ band as the fiducial, all other reddenings for each of the colors are coupled. The reddening equations than be re-framed as a linear algebra problem and solved for $A_\text{G}$:



\begin{subequations}
\label{equ:ag}
    \begin{align}
        A_G &= \frac{\mathbf{A}^\prime (\lambda) \cdot \mathbf{E}(\lambda)}{|\mathbf{A}^\prime (\lambda)|^2}\\
        \mathbf{A}^\prime (\lambda) &= 
            \begin{cases}
                \frac{A_{BP}}{A_G} - 1 & \text{if } \lambda = {BP}\\
                1 - \frac{A_{\lambda}}{A_G} & \text{if } \lambda \neq {BP}
            \end{cases}\\
        \mathbf{E}(\lambda) &= 
            \begin{cases}
                E(BP - G) & \text{if } \lambda = {BP}\\
                E(G - \lambda) & \text{if } \lambda \neq {BP}
            \end{cases}
    \end{align}
\end{subequations}

\noindent
where the $\frac{A_\lambda}{A_{\rm G}}$ can be found using an extinction law such as CCM89, and the components of $\mathbf{E}(\lambda)$ are the reddenings for each color.

\begin{figure}
    \centering
    \includegraphics[width=0.45\textwidth]{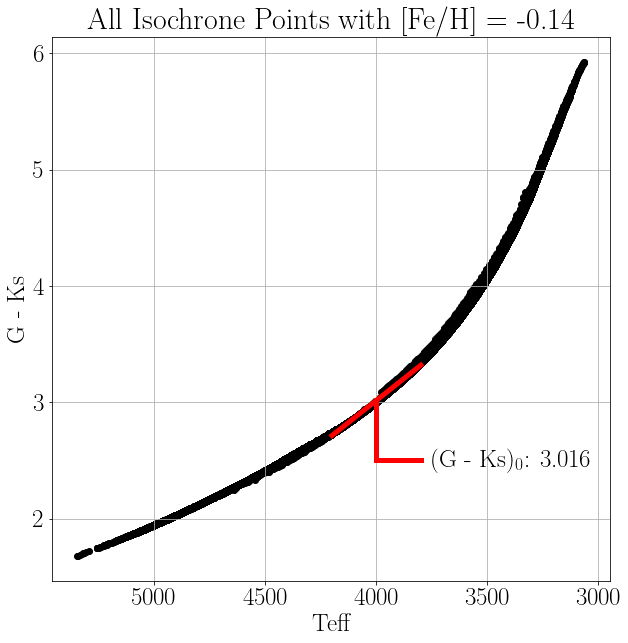}
    \caption{An example of how the intrinsic colors are determined from the isochrone photometry--\teff relations.}
    \label{fig:ages_extsch}
\end{figure}

The intrinsic color of a star can be found by interpolating isochrones with the following process:

\begin{enumerate}
    \item Pick isochrones with the Salaris corrected [Fe/H] of the star.
    \item Select all isochrone points with a temperature  within 200 K of the star's \teffe.
    \item The isochrone colors for ${BP} - G$, $G - {RP}$, $G - J$, $G - H$, and $G - K_{\rm s}$ are interpolated as functions of temperature using a B-spline. 
    \item Plugging in the \teff value of the star gives the expected intrinsic value for each of four colors previously mentioned and reddening is found. A pictorial example of calculating the $G - K_{\rm s}$ intrinsic color can be seen in Figure \ref{fig:ages_extsch}.
\end{enumerate}


Once $A_{\rm G}$ is calculated for a star, the chosen extinction law is used to calculate the extinction for all the other bands. The age of a star has a negligible effect on the color--\teff relationship, therefore, it is not considered when determining the extinction.

\subsection{Age Calculation}
\label{ssec:ages_agemethod}






The age of a star can be determined by comparing the observed photometric and spectroscopic parameters to stellar isochrones.  Specifically, the absolute multiband photometry is the most sensitive parameter to age.  Given a star's \teff and Salaris-corrected [Fe/H], we compare the observed multiband absolute magnitudes and surface gravity to the isochrones to determine age.

\begin{figure*}
    \centering
    \includegraphics[width=\textwidth]{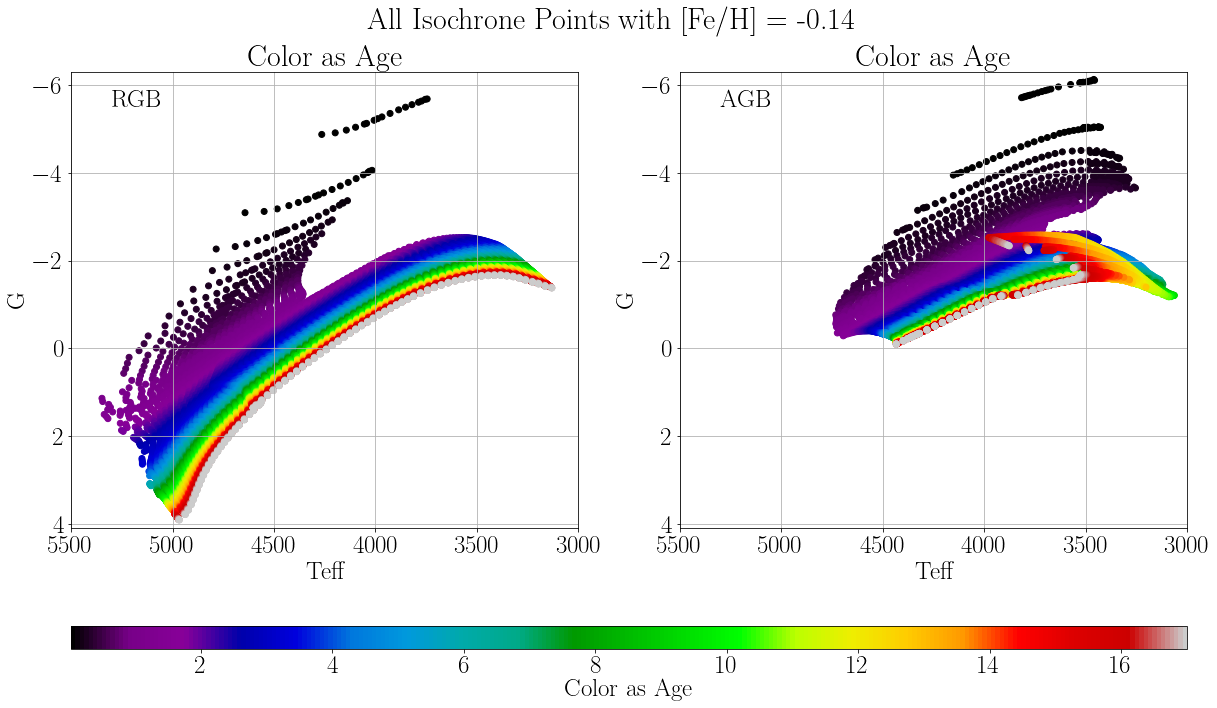}
    \caption{({\em Left}) The $G$--\teff relationship for [Fe/H] $=-0.14$ for the RGB isochrone points. There is a lot of degeneracy in the magnitude--\teff relation, but clearly broken with age. This is the basis for using the interpolations in section \ref{ssec:ages_agemethod} to get the ages. ({\em Right}) The same as the left panel, but for the AGB isochrone points. The AGB points mostly cover the gap in the RGB isochrone points between the young branch the rest of the points.}
    \label{fig:ages_mag_teff}
\end{figure*}

\subsubsection{Interpolated Isochrone Parameters}
\label{sssec:ages_interpmaglogg}






Within our isochrone grid, which is finely-sampled in [Fe/H] and age, we must interpolate a number of properties both ``along'' the isochrone (i.e., in mass) or between the grid points (in [Fe/H] and age).  The PARSEC isochrones provide an \texttt{int\_IMF}\footnote{Previously called \texttt{FLUM}.} column which is the indefinite integral of the IMF (initial mass function) by number from zero to the current mass.  The difference in \texttt{int\_IMF} between two neighboring isochrone points ($\Delta$\texttt{int\_IMF}) provides the number of stars occupying that isochrone segment (or stellar mass) per unit mass of the total initial stellar population.  For our purposes, $\Delta$\texttt{int\_IMF} essentially gives the expected number of RGB versus AGB stars or their relative probability. More information about why $\Delta$\texttt{int\_IMF} is important see section \ref{ssec:ages_weighting}
The isochrone absolute magnitudes, \logge, and $\Delta$\texttt{int\_IMF}
for a given age are interpolated through the following process:

\begin{enumerate}
    \item Pick all isochrone points with the Salaris-corrected [Fe/H] of the star regardless of age.
    \item Determine an initial guess for the age by calculating $\chi^2$ using ${BP}$, $G$, ${RP}$, $J$, $H$, $K_{\rm s}$, \teffe, and \logg (see Eq.\ \ref{equ:magchi}) for all selected isochrone points and picking the one with the lowest $\chi^2$ value.
    \item Determine the closest two adjacent ages of the initial guess age. If no isochrone points exist for the chosen age and [Fe/H], then pick the next closest age.
    \item Remove all isochrone points outside a specified \teff range as these points should not affect the interpolations. For most stars this amount to keeping isochrone points within 200 K of the measured \teff of the star.
    \item For each of the two ages the \teffe--magnitude relationship for each band, \teffe--\logge, and \teffe--$\Delta$\texttt{int\_IMF} relationships are used to interpolate these quantities for the star's observed \teffe.
    \item Finally, the two sets of interpolated isochrone quantities are linearly interpolated in age for the desired age.
\end{enumerate}

Normally, the \teffe--magnitude relationship for any given [Fe/H] exhibits a large degeneracy. However, this degeneracy can be broken with age as seen in Figure \ref{fig:ages_mag_teff}. This work exploits this to obtain the age of a star because the photometry and \teff are known. Also the boundary between the RGB and AGB phases can be blurred due to uncertainty in the \teff or magnitude values of a star as well as the close proximity of RGB and AGB isochrone points. The overlap is illustrated in Figure \ref{fig:ages_mag_teff}. This means that it is possible either RGB or AGB isochrones points could potentially describe a star due to overlap and the true evolutionary phase of a star is uncertain (see Figure \ref{fig:ages_mag_teff}).


If the age is considered a free parameter, then it can be found by fitting the isochrone absolute magnitudes, and \logg and matching to the observed absolute magnitudes and surface gravity. To accomplish this we make use of \texttt{scipy.optimize.curve$\_$fit} (hereafter \texttt{curve\_fit}). We limit the maximum number of iterations in \texttt{curve\_fit} to 5000.


The best age is determined by looking at the $\chi^2$ value for the final photometry and \logg determined by \texttt{curve\_fit}. Here the $\chi^2$ value is calculated by 





\begin{equation}
\label{equ:magchi}
    \chi^2 = \sum_\lambda \bigg(\frac{m_{\lambda,\text{iso}}-m_{\lambda,\text{obs}}}{\sigma_{\lambda,\text{obs}}}\bigg)^2 + \bigg(\frac{\log{g}_\text{iso}-\log{g}_\text{obs}}{\sigma_{\log{g},\text{obs}}}\bigg)^2,
\end{equation}

\noindent
where $m_{\lambda,\text{iso}}$ is the isochrone apparent magnitude for the $\lambda$ band (i.e., $BP$, $G$ $RP$, $J$, $H$ or $K_{\rm s}$), $m_{\lambda,\text{obs}}$ is the observed apparent magnitude, and $\sigma_{\lambda,\text{obs}}$ is the measured uncertainty in the observed $\lambda$ band magnitude. The second term is the \logg contribution to the $\chi^2$ where $\log{g}_\text{iso}$ is the value predicted from the isochrone interpolation, $\log{g}_\text{obs}$ is the observed surface gravity, and $\sigma_{\log{g},\text{obs}}$ is the uncertainty in the observed value of the surface gravity. The additional \logg term in $\chi^2$ produces improved fits and uses all of the observed information.



\subsection{Mass Calculation}
\label{ssec:ages_massmethod}

The mass and age of a star are highly anti-correlated on the RGB, which means it is generally straightforward to calculate the mass of a star if its age is known and vice versa. There is a slight degeneracy in the age-mass relationship that is broken with [Fe/H], which is a known quantity for our stars. At an age of 10 Gyr the isochrones mass can vary up to about $\pm$ 0.28 M$_\odot$ and at an age of 1 Gyr the isochrones mass can vary up to about $\pm$ 0.33 M$_\odot$.
To determine the mass for a star, it is interpolated as a function of age for a given Salaris-corrected [Fe/H].



\subsection{Evolutionary Phase Weighting}
\label{ssec:ages_weighting}

While most of the stars in our sample are RGB stars, there should also be a small amount of ``contamination'' from other phases of evolution such as the AGB. The tip of the APOGEE RGB (TRGB) selection is at a magnitude of $H \sim$12.35, but there can be overlap between the AGB and RGB regions in color magnitude space close to this magnitude \citep{nidever20lazy}.
Figure \ref{fig:ages_mag_teff} shows [Fe/H]=$-$0.14 isochrones in the $G$--\teff plane for a range of ages.  Only the RGB phase is displayed in the left panel which shows a sizeable gap between the older ages (> 1.0 Gyr) and very young ages ($\lesssim$ 1 Gyr).  There are stars in our APOGEE sample that fall in that gap.  As can be seen in the right panel, this gap is ``filled'' with the AGB phase with some overlap with the RGB.
Since there is no easy way to separate the RGB and AGB stars in our sample, 
we take a weighted sum between the ages calculated for RGB and AGB points.

This phase weighting is done through the following process:

\begin{enumerate}
    \item The age and mass for each star is calculated for each phase (RGB and AGB) separately without allowing for extrapolating outside the specified \teff range as described in Sections \ref{ssec:ages_agemethod} and \ref{ssec:ages_massmethod}.
    \item If a star is covered by only one phase, then that age and mass are used. If a star is covered by both phases, then the mean of the RGB and AGB mass and age values are calculated weighted by each phase's $\Delta$\texttt{int\_IMF}.
    \item If neither phase covers the star within the specified \teff range, then the code is allowed to extrapolate up to a limit of 200 K.  A star is assigned a ``bad'' value if it is beyond this 200 K extrapolation limit.
\end{enumerate}

$\Delta$\texttt{int\_IMF} represents the number of stars in a 1 solar mass total stellar population between the two points of the integrated IMF (\texttt{int\_IMF}).  This can be thought of as a the probability of detecting a star in this phase.

The \texttt{int\_IMF} is provided as part of the isochrone table from PARSEC and we use the Kroupa IMF \citep{kroupa2001imf,kroupa2002imf}. The $\Delta$\texttt{int\_IMF} is chosen for the weighting factor as it naturally encodes the fact that there should be more RGB stars compared to AGB stars.

\section{APOKASC Validation}
\label{sec:ages_kasccalval}

We use the APOKASC dataset to validate and calibrate our ages since it has highly accurate asteroseismic ages and uses an independent technique. The method used by APOKASC to derive ages relies on calculating the mass using temperature and asteroseismic scaling relations to obtain a measured surface gravity with APOGEE [M/H] and [$\alpha$/M]. An evolutionary track is then used for the corresponding mass, [M/H] and [$\alpha$/M]. The age of a star is then found by determining the point on the evolutionary track that has a \logg value equal to the asteroseismic \logge. While the APOKASC method does rely on measured APOGEE DR16 \citep{ahumada2020sixteenth} values of \teffe, [M/H], and [$\alpha$/M], the rest of the method is completely independent. This method applies to the APOKASC 3 ages in the \texttt{A3P\_AGEMOD\_JT} column. For more on previously calculated ages included with the APOKASC catalog, which may rely on slightly different methods, please see \cite{serenelli2017first} or \cite{pinsonneault2018apokasc}.

\subsection{APOKASC Extinctions}
\label{ssec:ages_kascext}



Accurate photometry is required to calculate the age of a star and, therefore, is important that the star's extinction is well-determined. The APOGEE catalog includes the \cite{schlegel1998reddening} $E(B - V)$ value for every star. We follow \cite{schlafly2010blue} in converting the Schlegel reddening values to extinctions for any band with a correction factor:

\begin{equation}
    \text{A}_{\rm \lambda} = 
    \begin{cases}
        0.78/1.32 \times E(B-V)_{\text{SFD}} \times \frac{\text{F99}(\lambda)}{\text{F99}(1 \mu m)} & \text{if } E(B - V) > 1\\
        0.78 \times E(B-V)_{\text{SFD}} & \text{if } E(B - V) < 1
    \end{cases}
\end{equation}

\noindent
where $E(B - V)_{\text{SFD}}$ is the Schlegel reddening value, F99($\lambda$) is the value of the \cite{fitzpatrick1999extinction} extinction law for the $\lambda$ band and F99($1 \mu m$) is the \cite{fitzpatrick1999extinction} extinction law value for a wavelength of 1 $\mu$m.

\begin{figure*}
    \centering
    \includegraphics[scale=0.25]{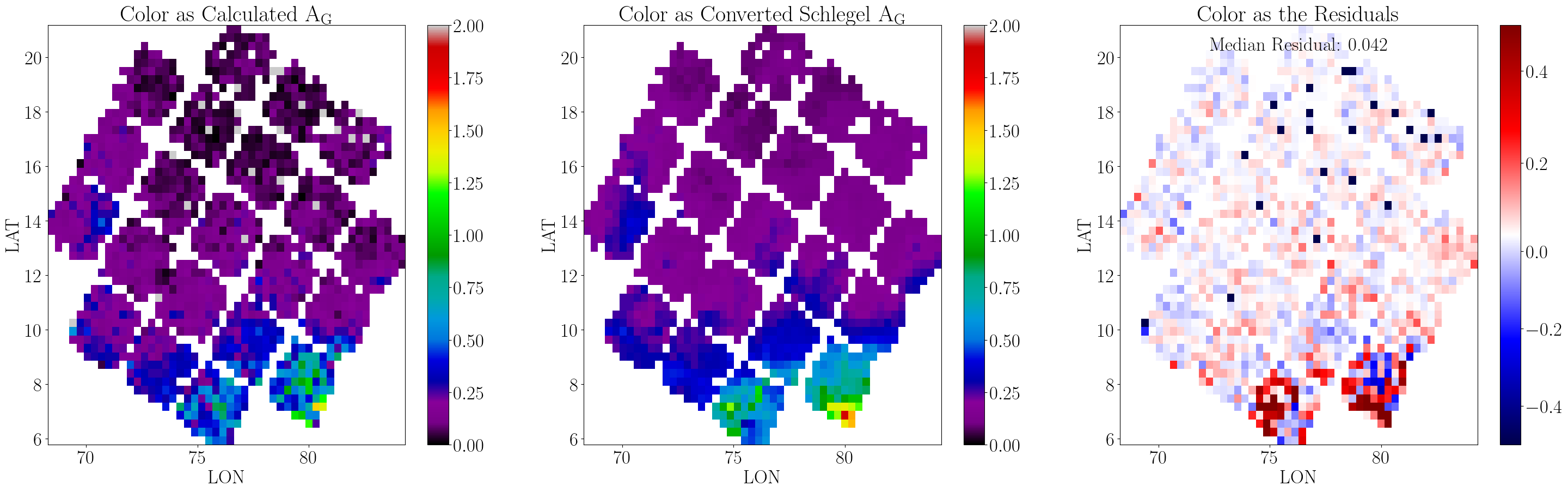}
    \caption{(Left) A$_{\rm G}$ extinction map converting the \protect\cite{schlegel1998reddening} reddening values using \protect\cite{schlafly2010blue} and \protect\cite{fitzpatrick1999extinction}. (Center) A$_{\rm G}$ map using the values calculating using the method in this work. (Right) The residuals between the Schlegel and calculated A$_{\rm G}$ values with a median residual of 0.043 mag.}
    \label{fig:ages_kasc_schlegel}
\end{figure*}

Comparing the A${\rm _G}$ expected from \cite{schlegel1998reddening} to the calculate A${\rm _G}$ in Figure \ref{fig:ages_kasc_schlegel} shows good agreement with a slight offset of about 0.043 mag. The part of the Kepler field nearest the MW plane tends to agree less well than the rest of the field.  However, this is a good validation of the extinction values derived from our method.

\begin{figure*}
    \centering
    \includegraphics[scale=0.245]{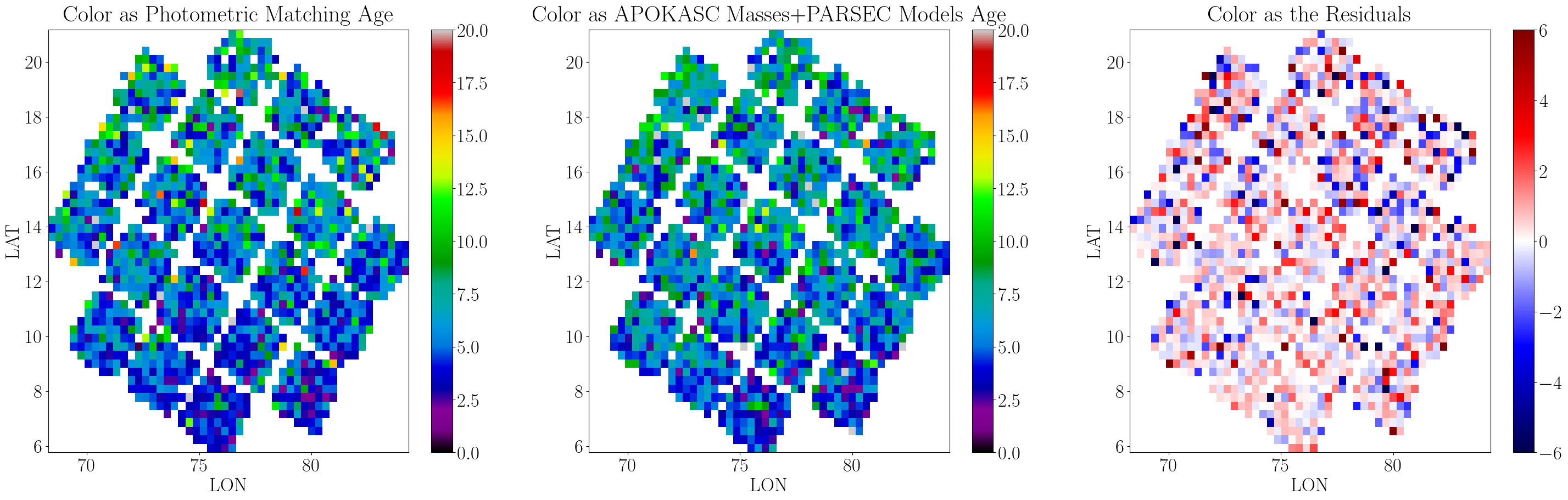}
    \caption{(Left) Age map of the Kepler field using APOKASC 3 masses and PARSEC isochrones. (Center) Age map of the Kepler field calculated using our method described in Section \ref{ssec:ages_agemethod}. (Right) The residuals between the methods with a median residual of 0.2 Gyr.}
    \label{fig:ages_kasc_agemap}
\end{figure*}

\subsection{APOKASC Ages}
\label{ssec:ages_kascage}

While our derived ages rely on using the PARSEC isochrones, the APOKASC method did not use PARSEC models in their calculations to convert mass to age.  This means that a direct comparison of our ages to the APOKASC catalog ages might have discrepancies because of the use of different models. Instead, we use the Salaris-corrected [Fe/H] and the APOKASC mass to determine the ages by interpolating the PARSEC isochrones. This is very similar to the mass calculation described in Section \ref{ssec:ages_massmethod}, but essentially in reverse.  From now on, we shall call these ages derived from the APOKASC mass and PARSEC isochrone the ``APOKASC ages''.


Comparing the APOKASC ages and our ages calculated through the photometric matching process described in Section \ref{ssec:ages_agemethod} shows quite good agreement. Figure \ref{fig:ages_kasc_agemap} shows the spatial map of the Kepler field ages from the two methods as well as the difference. Visually, the maps look very similar and the right panel showing that the residuals have an average close to zero, as desired. 

\begin{figure}
    \centering
    \includegraphics[scale=0.33]{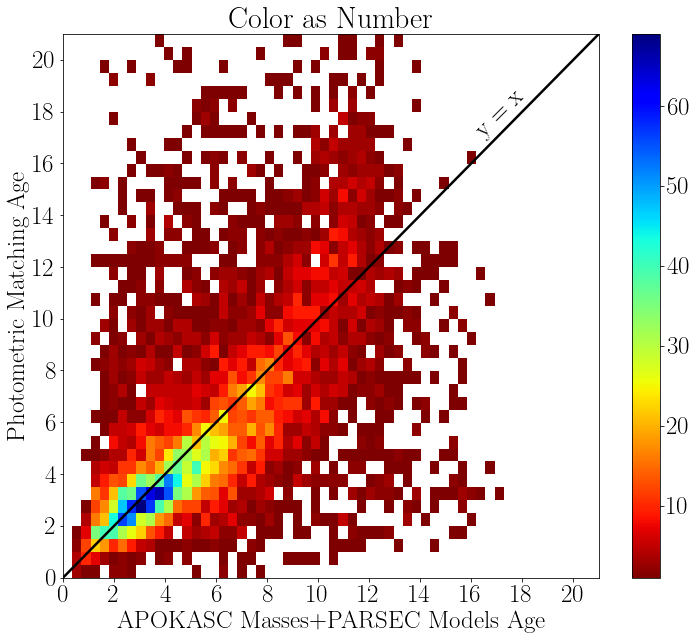}
    \caption{An age-to-age plot comparing the ages derived using the APOKASC masses and PARSEC models and ages derived using the photometric matching described in section \ref{ssec:ages_agemethod} for the APOKASC validation set colored by the number of stars in each bin. To help guide the eye the 1-to-1 line has been overplotted in black. It is clear especially for younger stars that the photometric matching ages tend to be slightly younger, though overall there is quite good agreement.}
    \label{fig:ages_ageagekasc}
\end{figure}

A direct one-to-one comparison of the APOKASC ages against our ages shows more clearly how well the two methods agree (see Figure \ref{fig:ages_ageagekasc}).  Overall, the locus of the stars in the 2D histogram is very close to the one-to-one line. Ages under 10 Gyr typically agree better and especially under $\sim$8 Gyr. This is not unexpected as older ages have larger uncertainties because the isochrones tend to stack up on each other in color--magnitude space. A curious feature is the upturn that happens for APOKASC ages older than about 10 Gyr. While we have investigated various causes for this systematic behavior, its origin remains unclear.  This is not as bad as it seems, because there are few stars older than 10 Gyr.  However, it does cause a larger spread of our oldest ages. 

As for systematic uncertainties, we find that the dispersion in the residuals is $\sim$1 Gyr for younger stars and then increase linearly up to about 3 Gyr for older stars close to the age of the universe.  This is a roughly $\sim$20\% age uncertainty.

\subsection{APOKASC Masses}
\label{ssec:ages_kascmass}

\begin{figure}
    \centering
    \includegraphics[scale=0.33]{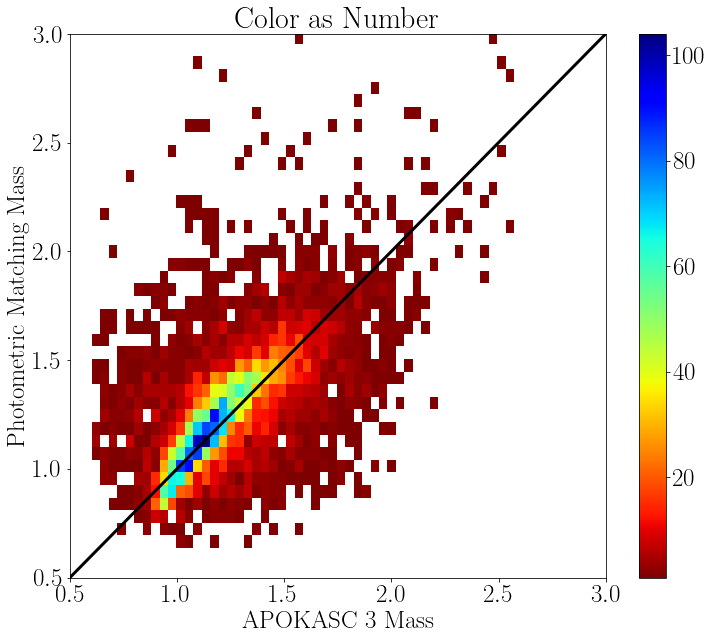}
    \caption{Comparing the APOKASC 3 masses to the masses calculated using Section \ref{ssec:ages_massmethod}.}
   \label{fig:ages_kasc_mass2mass}
\end{figure}

The mass and age of a star can be related through the mass-age relation. As another check this relation can be use to validate the ages. If the ages are correct then the masses calculated should match closely to the APOKASC catalog mass. After calculating the masses using section \ref{ssec:ages_massmethod}, the mass-to-mass plot in Figure \ref{fig:ages_kasc_mass2mass} shows good agreement. The locus of the stars in that plot match very well with the 1-to-1 black line. The number of stars in each bin quickly fall off moving from the line. The derived mass calculated from the photometric matching method appear to be slightly higher than APOKASC 3 values. Nonetheless, this is a reassuring result and suggests that the age calculation method in section \ref{sec:ages_extagemass} works well.


\section{Bias Correction}
\label{sec:ages_biascorr}

There are a number of biases in our dataset that need to be corrected for to obtain reliable age-related measurements.  There are two main biases that we correct for: (1) the targeting selection function, and (2) the number of expected RGB and AGB stars in our target selection box as a function of age and metallicity.

In general, one is not able to spectroscopically target and observe all of the stars in a given target selection category.  This was also the case for the APOGEE-2S Magellanic Cloud fields.  Our main goal was to obtain an integrated S/N=100 (over all visits) for a given star.  The LMC fields were allotted nine $\times$ $\sim$1 hour visits which meant that the S/N=100 could be achieved for H$\leq$12.8 stars.  Therefore, the main RGB targeting category (\texttt{BrtRGB}) was from $H$=12.0, the nominal tip of the RGB, to $H$=12.8.  In the inner fields, there were many more \texttt{BrtRGB} targets than could be accommodated by the APOGEE spectrograph ($\sim$250 science targets per plate).  This required us to select a subset of the targets which was accomplished by randomly drawing 260 stars (allowing for a small buffer to account for fiber ``collisions'').
%
In the outer fields, with lower stellar density, there were often less than 260 targets in the {\tt BrtRGB} box, and, therefore, fainter targets ({\tt FntRGB}) were included to fill the shortfall.  The faint limit was extended only as faint as was needed to fill the 260 number of targets.  Therefore, the faint limit in the outer fields varies from field to field and extends in some cases to $H$=15.3.  For each field we calculate an average ``selection function'' which is calculated as
\begin{equation}
    \text{Selection~Function} = \frac{N_{\rm potential~targets}} {N_{\rm observed~targets}}
\end{equation}
For consistency and simplicity across all fields, we use $N_{\rm potential~targets}$ for all targets down to $H$=15.3.  The outer fields have selection function values near unity, while the inner fields have very large values indicating that only a small fraction of the potential targets were observed.

\begin{figure*}
    \centering
    \includegraphics[scale=0.27]{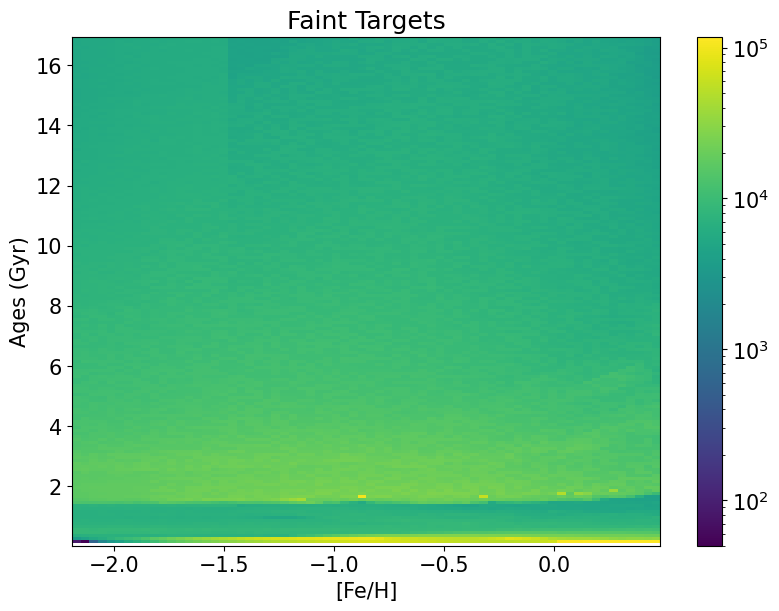}
    \includegraphics[scale=0.27]{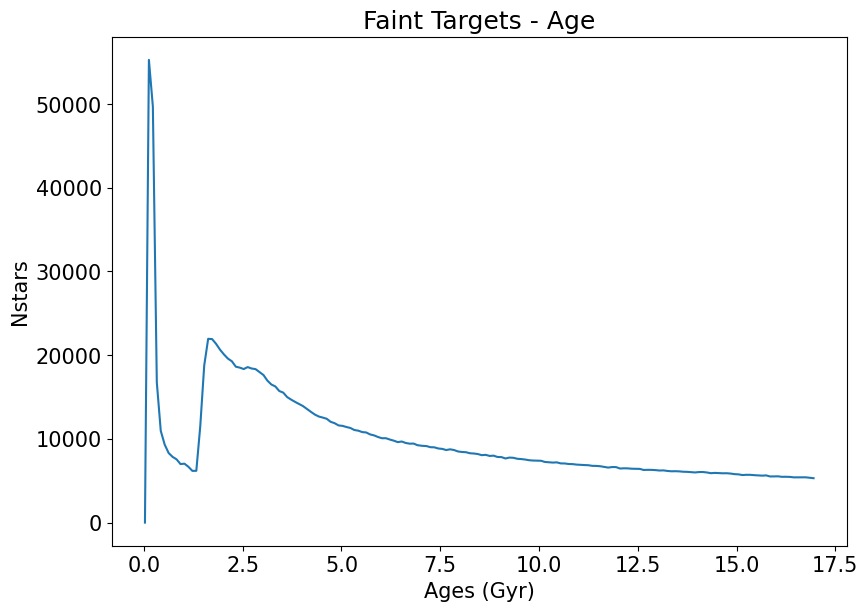}
    \includegraphics[scale=0.27]{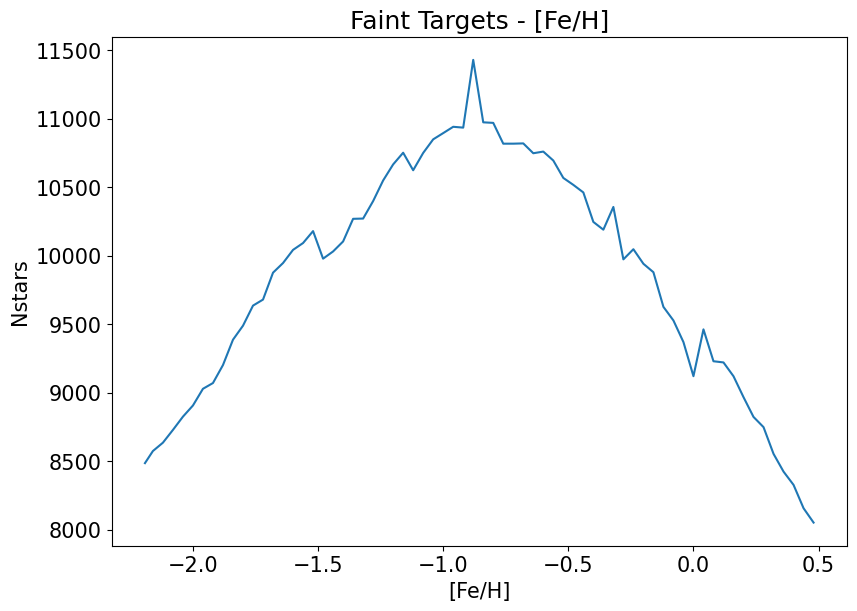}
    \caption{({\em Left}) Number of stars in the  target selection box for a stellar population of 10$^8$ \msun, as a function of age and metallicity.  There are regions of parameter space (e.g., old, metal-poor), where no stars are expected to be observed. 
    ({\em Middle}). Number of stars in the target selection box for a stellar population of 10$^8$ \msun, averaged over metallicity and showing the age dependence.  There is a sharp peak around 0.15 Gyr, and a ``valley'' at 0.3--1.5 Gyr.
    ({\em Right}) The same, but averaged over age and showing the metallicity dependence.}
    \label{fig:ages_synthphot}
\end{figure*}

From stellar evolution models we know that the number of RGB stars in a given magnitude range of a stellar population is not constant but varies systematically with age and metallicity.  This adds a separate ``astrophysical'' bias, which is especially important when using age histograms, because variations are expected even if the star formation history is constant.  We correct for this bias by calculating the number of stars expected in our target selection boxes for a given age and metallicity, $N_{\rm stars}$([Fe/H],Age), using synthetic photometry generated from PARSEC isochrones.  We use a total stellar population of 10$^8$ \msun and the RGB, horizontal branch, and AGB evolution stages.  As described above, the PARSEC $\Delta$\texttt{INT\_IMF} value is used to ascertain the number of synthetic stars to produce along the isochrone.  The left panel of Figure \ref{fig:ages_synthphot} shows the number of stars falling in the target selection box
as a function of metallicity and age
while the middle and right panels show average values for age and metallicity, respectively.  A substantial amount of structure is seen for younger ages, with a peak at $\sim$0.15 Gyr and a ``valley'' for 0.3--1.5 Gyr.  These are significant biases that need to be accounted for.

For a given star, we want to determine the amount of stellar population mass ($M_{\rm SP}$) that this star ``represents''. We convert our number of stars in the target selection box as a function of age and metallicity, $N_{\rm stars}$([Fe/H],Age), into a stellar population mass by taking the total stellar mass (10$^8$ \msune) used to generate our synthetic photometry (described in the previous paragraph) and dividing by the expected number of stars in the target selection box:
\begin{equation}
M_{\rm SP}\text{([Fe/H],Age)} = \frac{10^8 M_{\odot}}{N_{\rm stars}\text{([Fe/H],Age)}}
\end{equation}
Note, that our synthetic photometry did not distinguish between RGB or AGB stars but just counted all stars falling into our target selection box.  This deals nicely with the evolutionary phase ambiguity mentioned above.

When determining $M_{\rm SP}$ for a given star, we take into account the uncertainty in its age and metallicity by using
a ``measurement'' probability distribution function ($P_{\rm meas}$([Fe/H],Age)) unique for each star and represented as a 2-D Gaussian:
\begin{equation}
    P_{\rm meas}\text{([Fe/H],Age)} = \mathcal{N}(\text{[Fe/H]}_0,\sigma_{\rm [Fe/H]}) \times 
    \mathcal{N}(\text{Age}_0,\sigma_{\rm Age}) 
\end{equation}
where $\mathcal{N}$() is the normalized Gaussian distribution.
In addition, we can use $N_{\rm stars}$([Fe/H],Age) itself as a probability distribution function, i.e. a prior.  Regions of higher values indicate that we are more likely to detect stars there than regions where the values are low or zero.  We convert our distribution of number of stars into a probability by simply normalized $N_{\rm stars}$([Fe/H],Age)
\begin{equation}
    P_{\rm stars}([Fe/H],Age) = \frac{N_{\rm stars}\text{([Fe/H],Age)}}{\sum N_{\rm stars}\text{([Fe/H],Age)}}
\end{equation}
We use this information to produce a ``joint'' PDF by taking the product of the two PDFs. 
\begin{equation}
    P_{\rm joint}\text{([Fe/H],Age)} = P_{\rm meas}\text{([Fe/H],Age)} \times P_{\rm stars}\text{([Fe/H],Age)}
\end{equation}
The final $M_{\rm SP}$ value for a given APOGEE star is then calculated by taking a weighted mean of $M_{\rm SP}$([Fe/H],Age) with the joint PDF.
Since the faint limit of the {\tt FntRGB} target selection box varies from field to field, we calculated the number of stars in the {\tt FntRGB} box for 0.1 mag intervals from 12.8 to 15.3.
Finally, the $M_{\rm SP}$ value is multiplied by the ``selection function'' (from the first step) to calculate the final selection function-corrected ``stellar population mass'' $M_{\rm SP,SF}$ value for one of our stars.  This is the value that is used for ``weighting'' in many of the following calculations.

\section{Results} 
\label{sec:ages_lmcresults}

    






\subsection{The LMC Extinction Map}
\label{ssec:ages_lmc_ext}

\begin{figure*}
    \centering
    \includegraphics[scale=0.200]{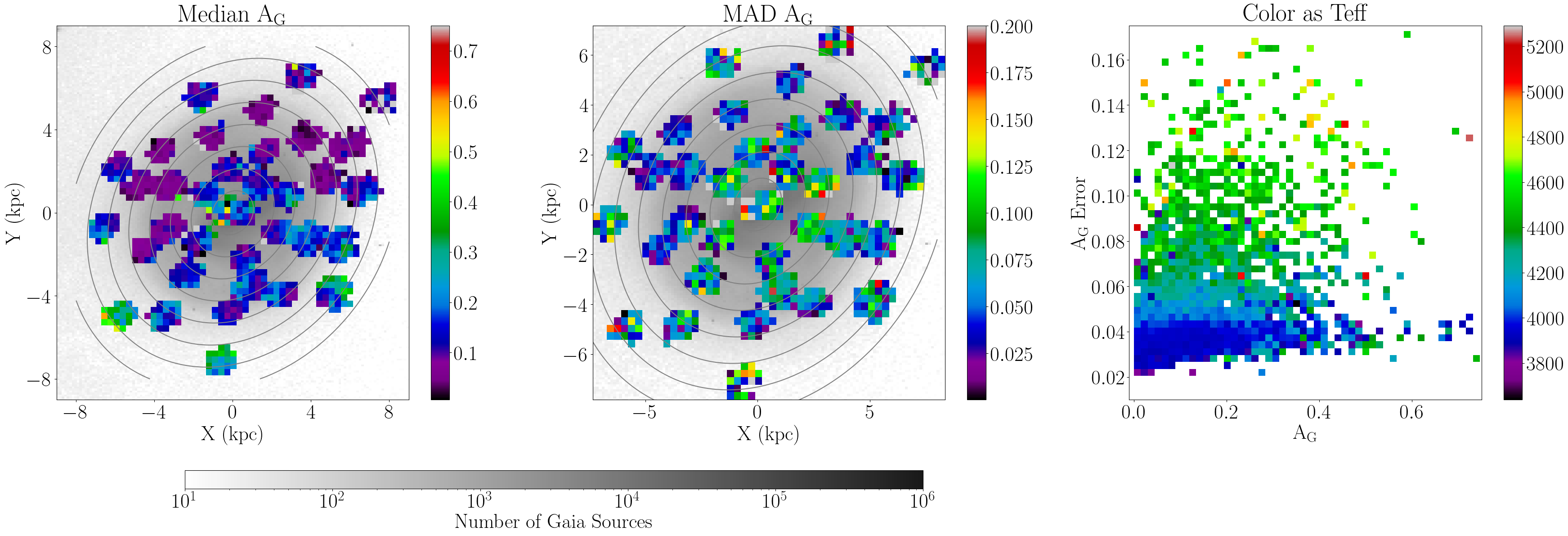}
    \caption{({\em Left}) The $G$ band extinction map for the LMC with the Gaia source number density in the background. In the map there is clearly a band of low extinction between $\sim$2 to $\sim$6 kpc in the north. (Center) The dispersion in extinction for each of the bins using MAD also with the Gaia background. ({\em Right}) The extinction errors as a function of the extinction colored by the \teff of each star. There is a definite correlation between the \teff of a star and calculate extinction error probably due to the steepness in the color--teff relations.}
    \label{fig:ages_lmc_map_ag}
\end{figure*}

The $G$ band extinction map for the LMC can be seen in Figure \ref{fig:ages_lmc_map_ag}. Interestingly there appears to be a band in the northern part of the LMC with low extinction for radii between $\sim$2 kpc and $\sim$6 kpc. This contrasts the southern part and center of the galaxy especially for the western side which is closest to the Small Magellanic Cloud (SMC).


\subsection{The LMC Age Map, Age-Radius Relation, \& Age Distribution}

\begin{figure*}
    \centering
    \includegraphics[scale=0.200]{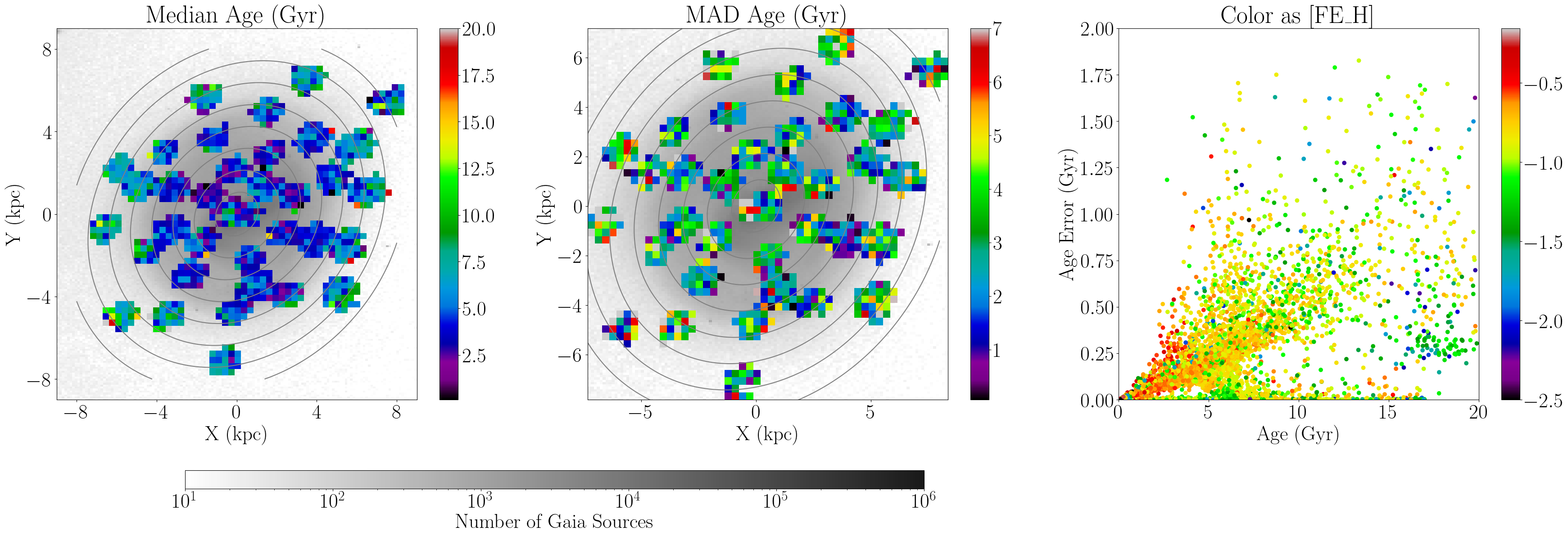} 
    \caption{({\em Left}) The age map for the LMC with the Gaia source number density in the background. ({\em Center}) The dispersion in age for each of the bins using MAD also with the Gaia background. ({\em Right}) The age errors as a function of the age colored by the [Fe/H] of each star. Most derived LMC have uncertainties that are $<$10\% suggesting great accuracy. There is a band of stars with uncertainties that appear to approach zero, but based on the visible envelop, the best course of action is to inflate any errors for these stars to at least 2.5\%.}
    \label{fig:ages_lmc_map_age}
\end{figure*}

The 2-D median age map of the LMC shows that the center of the galaxy tends to be slightly younger than the outskirts with a wide range of values for the age dispersion
(see Figure \ref{fig:ages_lmc_map_age}). A clear trend is evident in the age uncertainties, where older ages typically have larger uncertainties. The age uncertainties are calculated using standard error propagation techniques and interpolated splines. The uncertainties shown in the left panel of the age map. It is obvious that these uncertainties suggest great accuracy in the ages. When validating ages, Figure \ref{fig:ages_ageagekasc} shows that most stars fall within 10\% of the 1-to-1 line reinforcing the accuracy in the LMC ages. But regardless the derived uncertainties for the LMC ages should be considered lower limits as they are statistical uncertainties and other systematic sources of uncertainties could contribute. Some uncertainties of the LMC stars appear to be very close to 0 Gyr. For these stars it is probably best to impose a minimum uncertainty of $\sim$2.5\%. This estimation is based on clear envelop that can be seen in the left panel of Figure \ref{fig:ages_lmc_map_age}.




\begin{figure*}
    \centering
    \includegraphics[scale=0.335]{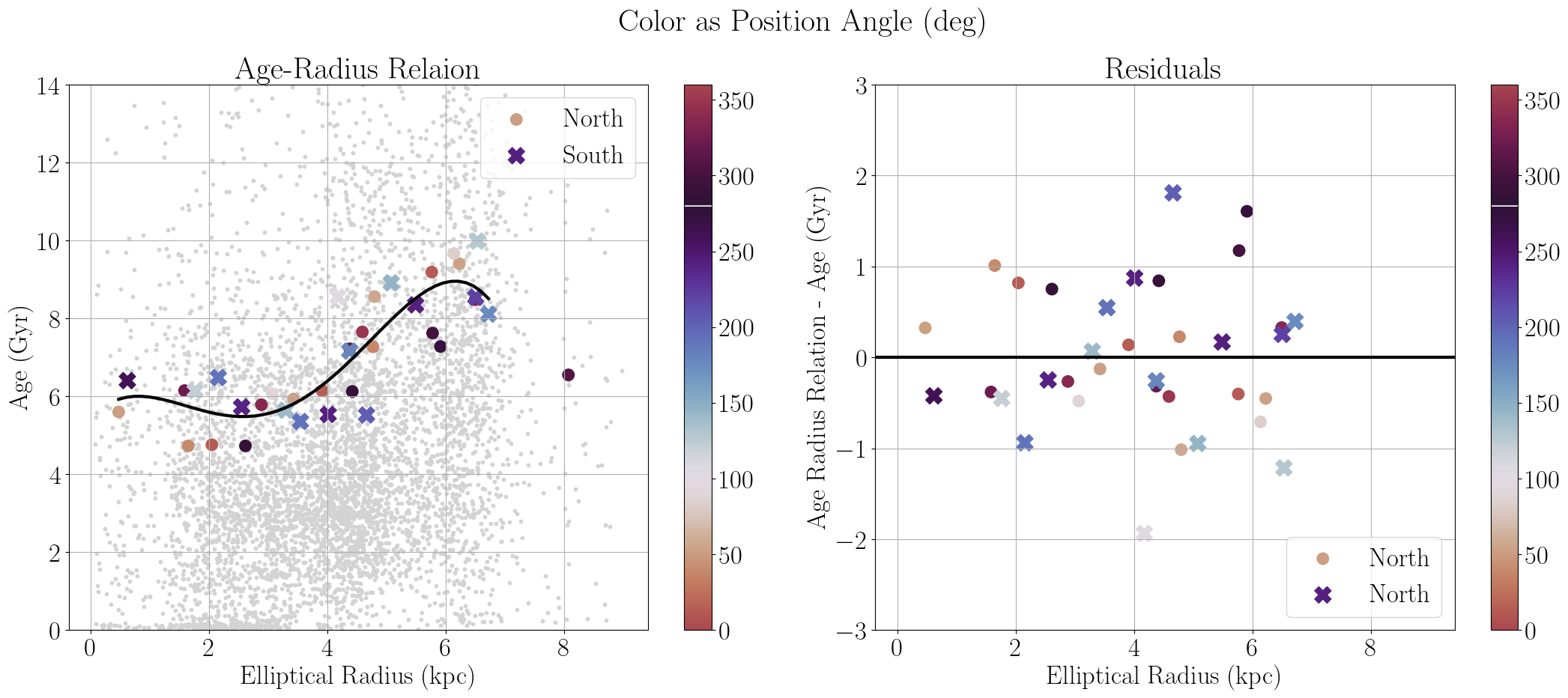} 
    \caption{({\em Left}) The age-radius relation for the LMC represented with the black solid line (see Equation \ref{equ:age_rad}). Individual LMC stars are shown in the background as light grey dots. The median ages for each of the LMC fields are overplotted as dots and xs colored in function of position angles. Northern fields are represented by dots while southern fields are represented by xs. It is clear that for inner radii that the northern fields are slightly younger than the southern fields at corresponding radii. ({\em Right}) The residuals of the quartic function fit. The MAD of the residuals is -0.19 for the clusters compared to 2.17 for the individual stars.}
    \label{fig:ages_lmc_age_rad}
\end{figure*}

To gain another perspective on the spatial distribution of stars, we look at the age-radius relation for the median age for each field. Fitting a quartic function to the median field ages gives:



\begin{equation}
    \label{equ:age_rad}
    {\rm age} = -0.03159r^4+0.4015r^3-1.443r^2+1.603r+5.452
\end{equation}



\noindent
where $r$ is the elliptical radius given by Equation \ref{equ:ellrad}. Quartic functions were chosen as visually the field points and the trends matched best without going to a higher order. A plot of the age-radius relations can be seen in Figure \ref{fig:ages_lmc_age_rad}. In the left panel, the nonlinearity is clearly evident. 



\begin{figure*}
    \centering
    \includegraphics[scale=0.6]{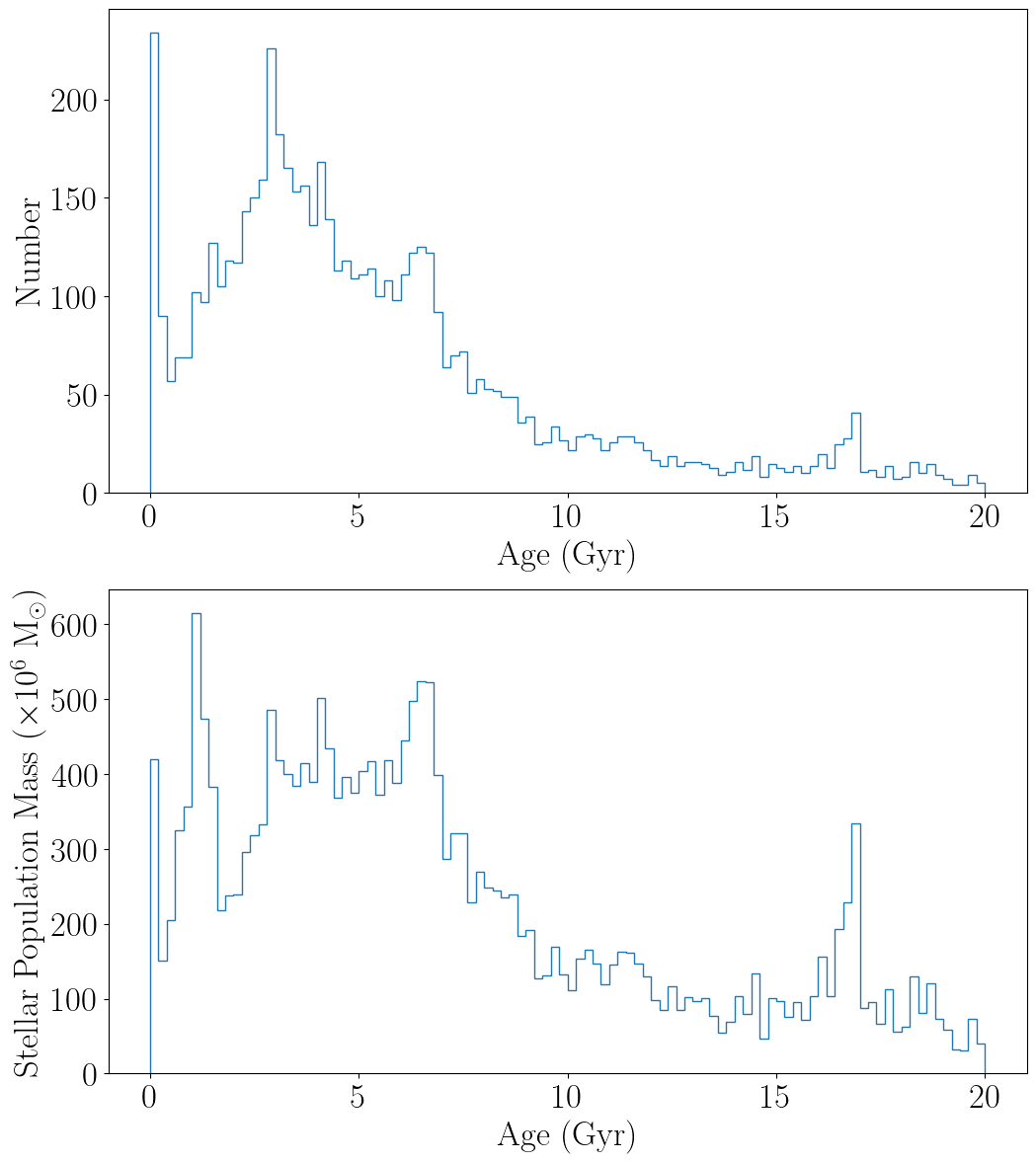}
    \caption{({\em Top}) A histogram of the ages calculated for the individual stars in the sample. The peak at 0 Gyr is made mostly of blue loop stars with some potentially bad results where stars could be assigned younger ages than they should be. The peaks less than 10 Gyr may suggest different star forming events while the peak at $\sim$17 corresponds to the age of the oldest isochrone. ({\em Bottom}) A histogram showing the stellar population mass corresponding to the calculated ages corrected for the bias according to section \ref{sec:ages_biascorr}. The first two peaks in the plateau between $\sim$2.5--5.0 Gyr do not appear to be temporally separated, but the third peak is. The supposed separation between the first two peaks may be an unphysical artefact.}
    \label{fig:ages_lmc_mass_age}
\end{figure*}

The age map suggests that young ages are widespread, but more centrally located.
The top panel of Figure \ref{fig:ages_lmc_mass_age} shows the age distribution of individual stars and indicates that there are substantially more young stars, though this could be misleading. Using the bias correction and stellar population mass calculation from Section \ref{sec:ages_biascorr} shows that the corrected-age distribution is somewhat different (see the bottom panel of Figure \ref{fig:ages_lmc_mass_age}). The first noticeable feature of the corrected distribution is the recent increase in stellar population mass (or star formation rate) consistent with many of the other results. Second, there is a wide peak or plateau for ages $\sim$3 Gyr to $\sim$7 Gyr.

\subsection{The LMC Age-Metallicity Relation}

\begin{figure*}
    \centering
    \includegraphics[scale=0.6]{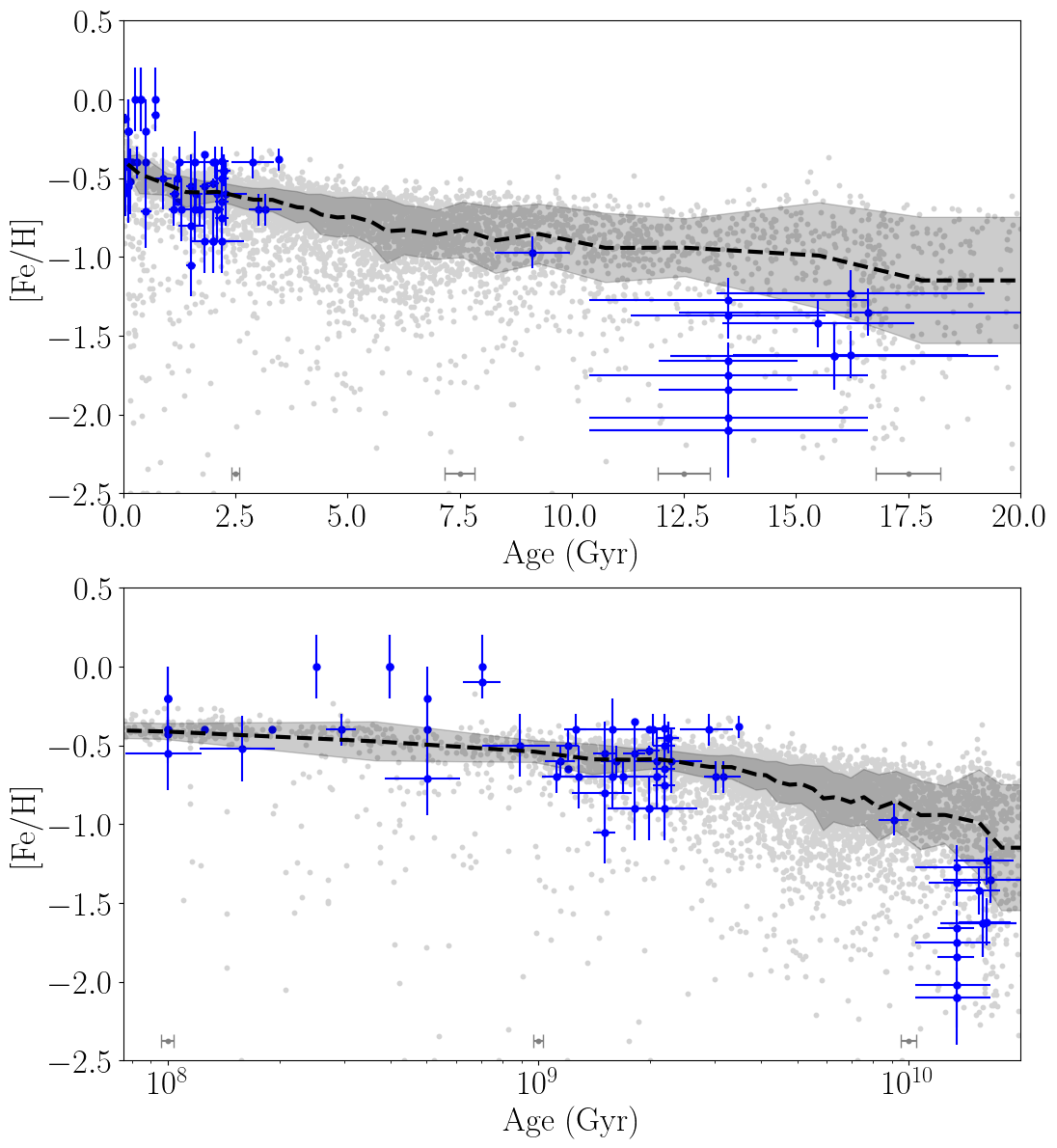} 
    \caption{({\em Top}) The AMR of the LMC calculated with the individual stars shown as the black dashed line. The APOGEE stars themselves are grey. The grey region shows the 1$\sigma$ limit around the AMR. The blue points with error bars represent the 85 LMC clusters from HZ09. The grey points at the bottom show the characteristic error in the grey points. Clearly the largest disagreement with clusters happens for the oldest ages where measurements get donimated by noise. ({\em Bottom}) The same as the top, but with a log scale which highlights younger ages.}
    \label{fig:ages_lmc_amr}
\end{figure*}

In Figure \ref{fig:ages_lmc_amr}, the age-metallicity relation (AMR) is fairly flat between 5 and 15 Gyr. Anything younger than 5 Gyr tends to show increasing metallicity with a rapid increase for stars with ages $\lesssim$2 Gyr, most likely related to the close interaction between the Clouds mentioned before. When comparing the AMR to LMC clusters from \cite{harris2009lmcsfh} (hereafter, HZ09), we find that the younger clusters tend to agree better than the older ones. Notably, the older clusters do bunch up at the age of the universe, but do not follow the AMR derived in this work.

\subsection{Asymmetric Drift}
\label{ssec:ages_asymmetric_drift}

\begin{figure*}
    \centering
    \includegraphics[scale=0.3]{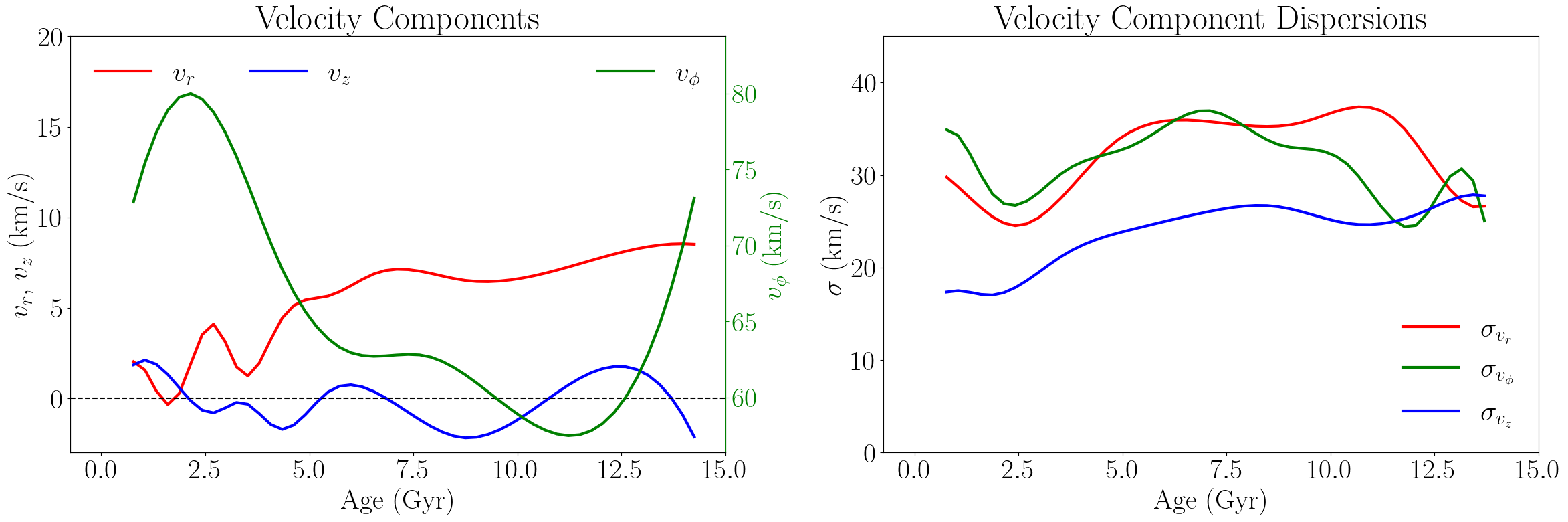}
    \caption{({\em Left}) The different components of the velocity as functions of age. The $v_r$ and $v_z$ are shown in red and blue respectively. These two velocity components use the black axis on the left. The 0.0 km/s for this is included as a dashed line. Both $v_r$ and $v_\phi$ components stay close to zero. The $v_r$ curve shows that there is an increase starting around ages of 5 Gyr, though for even the oldest ages this component does not get above 10 km/s. The $v_\phi$ velocity is shown as a green curve and uses the green axis on the right. For ages $\lesssim$2.5 Gyr there is a large increase in $v_\phi$ and then from 2.5--5.0 Gyr there is a decrease down to about $\sim$67 km/s and then an increase for the oldest stars. ({\em Right}) The different components of the velocity dispersion as functions of age.}
    \label{fig:ages_lm}
\end{figure*}

We briefly investigated the LMC's kinematics as a function of our derived ages by assuming that all of our stars lie in the LMC disk plane, and applying the disk kinematic and geometric model from \citet{Choi2022} to derive the velocity components $v_\phi$, $v_r$, $v_z$, and their associated errors, from our measured proper motions and line-of-sight velocities.  Figure \ref{fig:ages_lm} shows the mean values of the velocity components and their dispersions as a function of age, where we have restricted the sample to include only those stars with age$<$15 Gyr, velocity errors $<$15 km s$^{-1}$ in $v_\phi$ and $v_r$, errors $<$7.5 km s$^{-1}$ in $v_z$, and in-plane radius $R>$3 kpc; the limit on the radius is to ensure that we sample only the flat portion of the rotation curve.

As seen in the figure, the mean $v_z$ is close to 0 at all ages, while its associated dispersion rises from $<$20 km s$^{-1}$ at the youngest ages to nearly 30 km s$^{-1}$ at 10 Gyr and beyond.  This is expected behavior for a population that is dynamically heated over time.  The mean $v_r$ is also small at all ages, but slowly increasing with larger age to a value of $\sim$8 km s$^{-1}$ beyond 10 Gyr.  The dispersion in $v_r$ also grows with age, from $\sim$25 km s$^{-1}$ at 2.5 Gyr to $\sim$35 km s$^{-1}$ at 10 Gyr.  As we also see some correlation between $v_r$ and azimuthal angle, the small but positive $v_r$ may be a sign that the disk inclination changes in the periphery, where older stars predominate.  Finally, the mean $v_\phi$ clearly decreases with age, from a peak of $\sim$80 km s$^{-1}$ at 2.5 Gyr to a minimum of $\sim$60 km s$^{-1}$ at 10 Gyr, with roughly constant dispersion at all ages.  The significant decrease in mean $v_\phi$ is evidence for asymmetric drift, in which stars which started on nearly circular orbits acquire random deflections over time, which when combined with a decreasing density distribution as a function of radius leads to a suppression of the maximum observed rotational velocity.  We can use this observation of asymmetric drift to estimate the circular velocity of the LMC out to the radius limit of our sample.  As shown in \citet{BT1987}, we expect the magnitude of the asymmetric drift, $v_c - \overline{v_\phi}$ to be proportional to the radial velocity dispersion $\sigma_r^2$, as the effect of asymmetric drift comes from the preferential selection of stars on the slower portions of their elliptical orbits  at any given radius.  Thus if we can measure asymmetric drift at two different epochs $t_1$ and $t_2$, and assume that the density distribution is the same for stars with those ages, we can solve for the circular velocity $v_c$:
\begin{equation}
    v_c = \frac{v_{\phi,1} - v_{\phi,2}\sigma_{r,1}^2/\sigma_{r,2}^2}{1 - \sigma_{r,1}^2/\sigma_{r,2}^2}
\end{equation}
where $v_{\phi,1}$, $v_{\phi,2}$, $\sigma_{r,1}$, and $\sigma_{r,2}$ refer to the mean rotational velocity and radial velocity dispersion at epochs 1 and 2.  Taking  $v_{\phi,1}$ = 80 km s$^{-1}$, $v_{\phi,2}$ = 60 km s$^{-1}$, $\sigma_{r,1}$ = 25 km s$^{-1}$, and $\sigma_{r,2}$ = 35 km s$^{-1}$, we find $v_c\sim$100 km s${^1}$.  As the outermost in-plane radius of our sample is 10 kpc, we can then conclude that the mass of the LMC enclosed within 10 kpc is 2.3$\times$10$^{10}$ M$_\odot$.  If the rotation curve remains flat out to 15 kpc, as concluded by \citet{besla2015orbits}, then the total mass of the LMC is at least 3.5$\times$10$^{10}$ M$_\odot$, in good agreement with the value determined therein.

\section{Discussion}
\label{sec:ages_discconc}





\begin{figure*}
    \centering
    \includegraphics[scale=0.245]{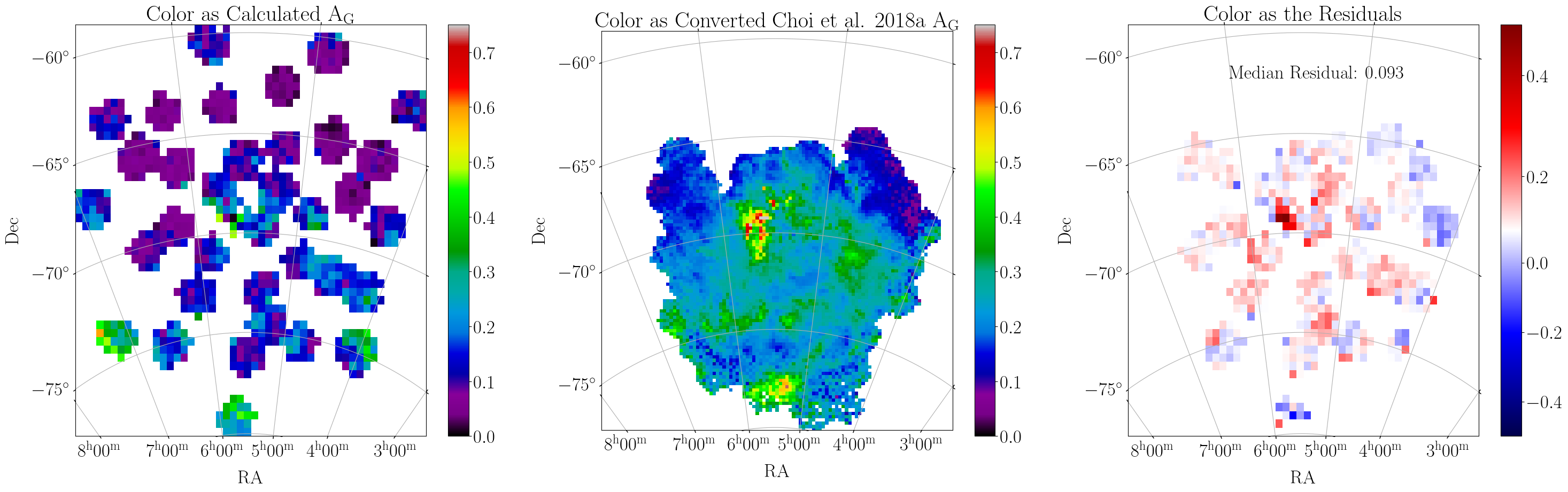}
    \caption{({\em Left}) The $G$ band extinction map for the LMC. This is the same map as seen in Figure \ref{fig:ages_lmc_map_ag}. ({\em Center}) The extinction map of LMC using A$_{\rm G}$ derived from \protect\cite{choi2018reddening} and the CCM89 extinction law. ({\em Right}) The residuals between the converted \protect\cite{choi2018reddening} A$_{\rm G}$ and the calculated A$_{\rm G}$ from this work with a median residual of 0.093 mag suggesting a slight offset.} 
    \label{fig:ages_lmc_ext_choi}
\end{figure*}

The extinction of the LMC has been recently explored with the SMASH data \citep{choi2018reddening}. In that paper, the extinction map was derived using red clump stars. This map was chosen to compare to because it is recent and has high resolution coverage of the LMC. In addition, the \cite{choi2018reddening} map agrees well with the average reddening value for the LMC found by \cite{zaritsky2004magellanic,haschke2011new,bell2022reddening} and also with the \cite{schlegel1998reddening} map. In order to compare the derived reddening and the extinction calculated in this work, CCM89 is used to convert the SMASH values to A$_{\rm G}$. In Figure \ref{fig:ages_lmc_ext_choi}, the left and right panels show the extinctions from this work and the SMASH $E(g - i)$ converted to A$_{\rm G}$, respectively. The right panel of the same figure shows the residuals which reveal a median offset of 0.093 mag while with the APOKASC validation there appears to be an offset of 0.043 mag suggesting that the method in this work slightly underestimates extinction compared to previous literature results. It is not entirely clear what the cause of this could be. The discrepancies in the extinction values may come from the spread of isochrone points in the isochrone photometry--\teff relations. The extinction method only fits to the locus of the isochrone points and the method in section \ref{ssec:ages_extmethod} essentially uses a weighted average of the across all bands, which may compound the width effect.

The age map for the LMC shows a large concentration of younger stars in the center of the galaxy with mean age increasing towards the outskirts. Many of the central young stars are thought to have been created as the result of an interaction between the LMC and SMC as the LMC experienced a burst of star formation at that time \citep{nidever20lazy}. This is one of the first times the spatial distribution of ages in the LMC has been done and so there are not many literature sources available to compare with.

The age-radius relation is consistent with this result showing a positive gradient (see the left panel of Figure \ref{fig:ages_lmc_age_rad}). The age-radius relation is clearly not flat for the LMC nor is it simply linear, but there is a distinct increase in the median age of stars with larger radii similar to what is seen in the age map. This shows that earlier in its history the LMC experienced more widespread star formation even out to its furthest limits, while at more recent times star formation has been concentrated toward the center, as previously mentioned.  This change in behavior could be due to the outer LMC gas being stripped off by the interaction of the LMC with the SMC or from the ram pressure in the Milky Way's hot halo. Another mechanism that can affect the distribution of stars in a galaxy is radial migration \citep[e.g.][]{sellwood2002radial,sellwood2014secular}. Radial migration leads to changes in the metallicity gradient of a galaxy and because metallicity is a proxy for age, this should also affect the age-radius relation.

In the left of Figure \ref{fig:ages_lmc_age_rad}, the majority of the northern fields are below the trendline for radii $<$ 4 kpc, but the southern fields tend to be above it producing a north-south dichotomy. There is a single spiral arm structure present in the inner northern fields with approximately the right age to be created through a known interaction between the LMC and SMC $\sim$2 Gyr ago. Through N-body simulations it has been shown that single spiral arms can be created through intergalactic tidal interactions \citep[e.g.][]{berentzen2003nbody,pearson2018nbody}. This suggests that the spiral arm in the north was likely created by this major event.





The AMR of the LMC is mostly flat, increasing very slowly over time. In the upper panel of Figure \ref{fig:ages_lmc_amr} there is a clear recent increase in [Fe/H] coincident with the LMC-SMC interaction. The derived AMR shows decent agreement for the LMC globular clusters from HZ09, except close to the age of the universe. This is also true of AMRs derived using clusters such as in \cite{olszewski1991spectroscopy}, \cite{dirsch2000age}, and \cite{grocholski2006ii}, which all exhibit the downturn near the age of the universe while the one derived in this work does not. The LMC AMR derived from individual RGB stars in \cite{carrera2008lmcchem} matches even better than the cluster AMRs, but the downturn is still seen. It appears that for field star derived AMRs, the metallicities for intermediate ages tends to be slightly higher than what is seen according to clusters. This could be due to a known lack of clusters for those ages \cite{rich2001new},\cite{bekki2004explaining}, and \cite{mackey2006photometry}. For the oldest ages, both the cluster and field star derived AMRs do show a downturn around the age of the universe that is not present in the AMR derived here. A possible reason for this is that the stars with ages $>$12.5 Gyr are dominated by measurement uncertainties in their parameters that get propagated through the age method. Other sources of this effect could be systematics with the PARSEC models or the interpolation favors pushing metal poor stars to older ages.


Previously the difference in the AMR for the LMC disk and bar have been investigated in \citep{cole2005amr,carrera2008lmcchem,carrera2011metallicities}. Selecting the APOGEE bar stars in this work, it was found that there is an increase in [Fe/H] in the last $\sim$1 Gyr for the both the bar and the LMC disk. This contradicts \cite{cole2005amr} and \cite{carrera2008lmcchem}, which found that there is not an appreciable increase in [Fe/H] for the bar in the last Gyr or so. Though there are less stars in the bar for older ages, the AMR for the bar is largely flat for ages older than 5 Gyr. For the most part the bar AMR suggests that these stars are more metal-rich for all times.

The detection of asymmetric drift in the stellar kinematics as a function of age is expected, and thus lends support to our ability to measure ages with our method.  We used the measured evolution in asymmetric drift to estimate the LMC's circular velocity and thus its total mass out to 10 kpc; our estimate of 2.3$\times$10$^{10}$ M$_\odot$ within 10 kpc is $\sim$10\% higher than e.g. \citet{vdM2014}, but in overall good agreement.

\section{Summary}
\label{sec:ages_summary}

We present an isochrone based method to calculate the ages of individual RGB stars and use this to find the ages of more than 6000 stars in the LMC. This method makes use of both photometry and spectroscopy and requires the distance to be known. The ``A codE to calculaTe stellAr ageS'' (Aetas) code used throughout this work is available online\footnote{\url{https://github.com/JoshuaPovick/aetas}}. In addition the age method can be easily adapted to find the age of any individual star under some minimal constraints. Aetas requires that the photometry, distance, [Fe/H], [$\alpha$/Fe], \teffe, and \logge be already measured. In order to use a different set of isochrones with Aetas, the appropriate labels will need to be changed to match the input table.

\begin{itemize}
    \item We have used accurate photometry, spectroscopy and isochrones to determine ages statistically accurate to $\sim$1--2\% and systematically accurate to $\sim$1--3 Gyr of stars in the LMC disk.
    \item Our method determined age and extinction simultaneously.  Our LMC extinction map corresponds well with other literature extinction maps \citep[e.g.,][]{choi2018reddening}.
    \item The median age is relatively flat with radius at $\sim$4 Gyr out to $\sim$4 kpc.  Then it rises to $\sim$6 Gyr until the edge of our coverage at 7 kpc.
    \item We see an Age dichotomy in the LMC disk, with northern fields slightly younger than same ones at the same elliptical radius. 
    \item Exploring the relation between the derived ages and kinematics of the LMC shows evidence of asymmetric drift.
\end{itemize}



Papers II and III in this series will use the ages determined in this work and study the chemical age-radius trends in both the LMC and SMC.

\section*{Acknowledgements}



J.T.P. acknowledges support for this research from the National Science Foundation (AST-1908331) and the Montana Space Grant Consortium Graduate Fellowship.
D.L.N. acknowledges support for this research from the National Science Foundation (AST-1908331)

Funding for the Sloan Digital Sky Survey IV has been provided by the Alfred P. Sloan Foundation, the U.S. Department of Energy Office of Science, and the Participating Institutions. 

SDSS-IV acknowledges support and resources from the Center for High Performance Computing  at the University of Utah. The SDSS website is www.sdss.org.

SDSS-IV is managed by the Astrophysical Research Consortium for the Participating Institutions of the SDSS Collaboration including the Brazilian Participation Group, the Carnegie Institution for Science, Carnegie Mellon University, Center for Astrophysics | Harvard \& Smithsonian, the Chilean Participation Group, the French Participation Group, Instituto de Astrof\'isica de Canarias, The Johns Hopkins University, Kavli Institute for the Physics and Mathematics of the Universe (IPMU) / University of Tokyo, the Korean Participation Group, Lawrence Berkeley National Laboratory, Leibniz Institut f\"ur Astrophysik Potsdam (AIP), Max-Planck-Institut f\"ur Astronomie (MPIA Heidelberg), Max-Planck-Institut f\"ur Astrophysik (MPA Garching), Max-Planck-Institut f\"ur Extraterrestrische Physik (MPE), National Astronomical Observatories of China, New Mexico State University, New York University, University of Notre Dame, Observat\'ario Nacional / MCTI, The Ohio State University, Pennsylvania State University, Shanghai Astronomical Observatory, United Kingdom Participation Group, Universidad Nacional Aut\'onoma de M\'exico, University of Arizona, University of Colorado Boulder, University of Oxford, University of Portsmouth, University of Utah, University of Virginia, University of Washington, University of Wisconsin, Vanderbilt University, and Yale University.

This work has made use of data from the European Space Agency (ESA) mission
{\it Gaia} (\url{https://www.cosmos.esa.int/gaia}), processed by the {\it Gaia}
Data Processing and Analysis Consortium (DPAC,
\url{https://www.cosmos.esa.int/web/gaia/dpac/consortium}). Funding for the DPAC
has been provided by national institutions, in particular the institutions
participating in the {\it Gaia} Multilateral Agreement.

\textit{Software:} Astropy \citep{pricewhelan2018astropy, robitaille2013astropy}, Matplotlib \citep{hunter2007matplotlib}, NumPy \citep{harris2020numpy}, SciPy \citep{virtanen2020scipy}

\section*{Data Availability}


All APOGEE DR17 data used in this study is publicly available and can be found at: \url{https://www.sdss4.org/dr17/data_access/}. This dataset does include the corresponding Gaia and 2MASS data in addition to the spectroscopic data.




\bibliographystyle{mnras}
\bibliography{bibliography} 




\appendix

\section{\textit{Gaia} DR3 selection}

For Figure \ref{fig:ages_map}, we use \textit{Gaia} DR3 data to make the background density map, which highlights some of the structures around the Clouds. We use a parallax cut of stars compatible with being more than 20 kpc away at lower than 10\% uncertainty. A proper motion cut of a radius of 1.05 mas/yr centered around $(\mu_{\mathrm{L}},\mu_{\mathrm{B}})=(1.767,0.451)$ in Magellanic Stream coordinates \citep{nidever08leadingarm}. A magnitude cut at G=19.3 and a cut around the red giant branch sequence in a colour-magnitude diagram.



\bsp	
\label{lastpage}
\end{document}